%% file: main.tex
\documentclass[
    aps,
    prx,
    reprint,
    superscriptaddress,
    longbibliography,
    floatfix,
]{revtex4-2}

\usepackage{amsmath}
\usepackage{mathtools}
\usepackage{braket}
\usepackage{footnote}
\usepackage{lipsum}
\makesavenoteenv{table}
\usepackage[version=4]{mhchem}
\usepackage{nicefrac}
\usepackage{orcidlink}
\usepackage{ragged2e}
\usepackage{placeins}
\usepackage{siunitx}
\usepackage{longtable}
\usepackage{orcidlink}
\usepackage{etoolbox}
\usepackage{graphicx}
\graphicspath{ {./figs/} }

\usepackage{hyperref}
\hypersetup{
    colorlinks=true,
    linkcolor=blue,
    urlcolor=magenta,      
    citecolor=cyan,
    }

\usepackage{float}

\newcommand{\refsub}[2]{\hyperref[#1]{\ref*{#1}(#2)}}
\newcommand{\s}[1]{\mathrm{#1}}

\newcommand{\valueC}{\ensuremath{C = \num{20.6 \pm1.1}}}
\newcommand{\valueCcoh}{\ensuremath{C_\s{coh} = \num{8.3 \pm 1.2}}}  

\newcommand{\valueDeltaT}{\ensuremath{\num{0.988 \pm 0.004}}}
\newcommand{\valueDeltaTPercentage}{\ensuremath{\num{98.8 \pm 0.4}}}
\newcommand{\valueNumQ}{327}

\newcommand{\valueQ}{\ensuremath{Q = \num{25.4 \pm 0.4}\times 10^3}}
\newcommand{\valueQavg}{\ensuremath{Q = \num{1 \pm 0.3}\times 10^4}}
\newcommand{\Qutech}{QuTech, Delft University of Technology, 2628 CJ Delft, The Netherlands}
\newcommand{\ICFO}{ICFO—Institut de Ciencies Fotoniques, The Barcelona Institute of Science and Technology,
Castelldefels (Barcelona) 08860, Spain}

\newcommand{\Kavli}{Kavli Institute of Nanoscience Delft, Delft University of Technology, 2628 CJ Delft, The Netherlands}
\newcommand{\Sandbox}{SandboxAQ, Palo Alto, California, USA}
\newcommand{\CNRS}{Univ. Grenoble Alpes, CNRS, Grenoble INP, Institut N ́eel, 38000 Grenoble, France}

\begin{document}

\title{
Above-Unity Coherent Cooperativity of Tin-Vacancy Centers in Diamond Photonic Crystal Cavities
}

\author{Nina Codreanu\orcidlink{0009-0006-6646-8396}}
\thanks{These authors contributed equally to this work.}
\affiliation{\Qutech}
\affiliation{\Kavli}

\author{Tim Turan\orcidlink{0009-0003-9908-7985}}
\thanks{These authors contributed equally to this work.}
\affiliation{\Qutech}
\affiliation{\Kavli}

\author{Daniel Bedialauneta Rodriguez\orcidlink{0009-0000-3634-070X}}
\thanks{These authors contributed equally to this work.}
\affiliation{\Qutech}
\affiliation{\Kavli}

\author{Matteo Pasini\orcidlink{0009-0005-1358-7896}}
\affiliation{\Qutech}
\affiliation{\Kavli}
\altaffiliation[Present address: ]{\ICFO}

\author{Lorenzo de Santis\orcidlink{0000-0003-0179-3412}}
\affiliation{\Qutech}
\affiliation{\Kavli}
\altaffiliation[Present address: ]{\CNRS}

\author{Maximilian Ruf\orcidlink{0000-0001-9116-6214}}
\affiliation{\Qutech}
\affiliation{\Kavli}
\altaffiliation[Present address: ]{\Sandbox}

\author{Christian F. Primavera\orcidlink{0000-0001-7486-6243}}
\affiliation{\Qutech}
\affiliation{\Kavli}

\author{Leonardo G. C. Wienhoven \orcidlink{https://orcid.org/0009-0009-7745-3765}}
\affiliation{\Qutech}
\affiliation{\Kavli}

\author{Caroline E. Smulders\orcidlink{0009-0006-2584-3946}}
\affiliation{\Qutech}
\affiliation{\Kavli}

\author{Simon Gröblacher\orcidlink{0000-0003-3932-7820}}
\affiliation{\Kavli}

\author{Ronald Hanson\orcidlink{0000-0001-8938-2137}}
\email{R.Hanson@tudelft.nl}
\affiliation{\Qutech}
\affiliation{\Kavli}
\date{\today}
% \affiliation{
% QuTech and Kavli Institute of Nanoscience, Delft University of Technology, PO Box 5046, 2600 GA Delft, The Netherlands

\begin{abstract}
    % \justifying
    \noindent 
    The tin-vacancy center in diamond (SnV) has emerged as a compelling building block for realizing next-generation quantum networks thanks to its excellent optical and spin properties. Coupling to photonic crystal cavities (PCCs) promises to further enhance the SnV light-matter interface and unlock a diverse range of entanglement generation protocols. Recent pioneering experiments showing Purcell enhancement of SnV centers in PCCs underscore this potential.
    However, optical coupling that is coherent - the key ingredient for use in quantum protocols -  has so far remained elusive.
    Here, we demonstrate above-unity coherent cooperativity of SnV centers embedded in photonic crystal cavities. 
    We fabricate free-standing PCCs using a quasi-isotropic undercut.
    Across two samples, we conduct room-temperature characterizations, measuring resonances for \valueNumQ{} cavities, with an average quality factor exceeding \valueQavg{}.
    Two cavity-coupled emitters are examined in detail, exhibiting quality factors up to \valueQ~and Purcell-reduced lifetimes corresponding to cooperativities up to \valueC. 
    Furthermore, the single SnVs are observed to strongly modulate the cavity transmission with an extinction contrast up to \valueDeltaTPercentage\% on resonance. 
    Finally, SnV linewidth measurements reveal above-unity coherent cooperativities in both devices, with the highest value being \valueCcoh.
    These results open the door to using cavity-coupled SnV centers as efficient, coherent light-matter interfaces for future quantum networks.
\end{abstract}
\maketitle

\section*{Introduction}
\begin{figure}[t]
    \centering
    \includegraphics[width=\linewidth]{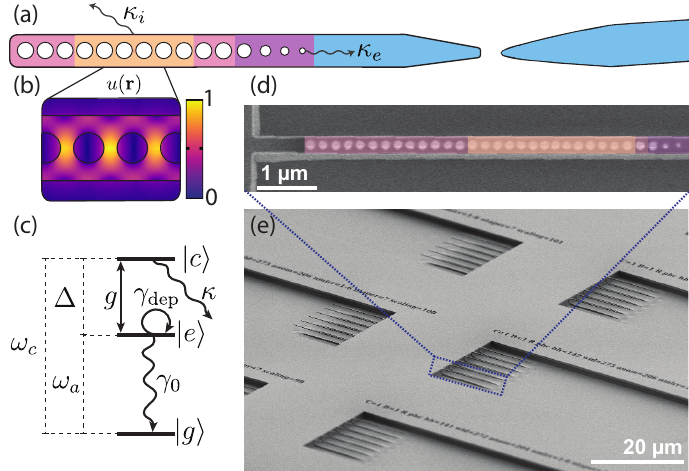}
    \caption{ Tin-vacancy centers in diamond photonic crystal cavities (PCC).
    \textbf{(a)} 
    Schematic of the device. 
    The PCC design is illustrated from left to right: strong mirror, defect region, weak mirror and taper end. We collect the light with a tapered optical fiber in a lensed configuration. 
    \textbf{(b)} Simulated electric field of the fundamental cavity mode normalized to the maximum
    $u(\mathbf r)$, see Appendix \ref{sec:cavity-simulations}.
    %The coupling strength $g$ between the SnV center and the cavity mode depends on its position within the cavity mode.
    \textbf{(c)}
    Simplified level structure of the combined emitter-cavity system. The emitter is described by a transition frequency $\omega_a$ between the ground state $|g\rangle$ and the excited state $|e\rangle$. The spontaneous emission rate is denoted $\gamma_0$ and $\gamma_\s{dep}$ is the dephasing rate. The state $|c\rangle$ denotes a photon in the cavity mode and the emitter in the ground state. The cavity mode has a tunable frequency $\omega_c=\omega_a+\Delta$ and its total decay rate is $\kappa =\kappa_i+\kappa_e$. The coupling strength between the cavity and the emitter is $g$.
    \textbf{(d)}
    SEM image of a free-standing PCC device, taken at a \SI{0}{\degree} angle.
    The device regions (false colored) correspond to the schematic in (a).
    \textbf{(e)}
    SEM image of the diamond sample taken at a \SI{65}{\degree} angle.
    It displays multiple arrays of cavities, where each array contains six PCC devices.
    }
    \label{fig:1}
\end{figure}
A future quantum internet, built using quantum processor nodes connected via optical channels, promises applications
such as secure communication, distributed quantum computation and enhanced sensing
\cite{kimble_quantum_2008,wehner_quantum_2018,gottesman_longerbaseline_2012}

Optically active spin qubits in diamond represent a promising building block, with recent demonstrations of multinode
quantum networks of remote solid-state qubits \cite{pompili_realization_2021,hermans_qubit_2022,wei_BQC,stolk_metropolitanscale_2024,knaut_entanglement_2024,sipahigil_integrated_2016}.
One approach to boost the rate of entanglement generation is the integration of diamond spin qubits into nanophotonic devices, such as waveguides and photonic crystal cavities (PCCs) 
\cite{parker_diamond_2024,pasini_nonlinear_2024,ding_highq_2024,rugar_quantum_2022,kuruma_coupling_2021,sipahigil_integrated_2016}
% ,ding_high-q_2024,stas_robust_2022,rugar_quantum_2022,kuruma_coupling_2021,sipahigil_integrated_2016}.
Optical cavities enhance light-matter interactions, increasing the probability of emission into the desired mode to
$\beta = C/(C+1)$, which approaches unity at high cooperativities $C$.
In particular, the small mode volumes of PCCs facilitate achieving high cooperativities, leading to high emission efficiencies
\cite{reiserer_cavitybased_2015} and opening the door to entanglement schemes in which the devices are operated as spin-photon gates \cite{beukers_remoteentanglement_2024}.

Tin-vacancy (SnV) centers in diamond have emerged as promising candidates to implement quantum networking hardware due to their favorable optical and spin properties. Recent experiments have demonstrated their suitability as quantum network nodes, including high-fidelity control of the electron spin
\cite{guo_microwavebased_2023,rosenthal_microwave_2023},
the use of nearby \ce{^{13}C} nuclear spins as memory qubits
\cite{C13_Hunger,beukers_control_2025},
hybrid integration with photonic chips for routing
\cite{wan_large-scale_2020,chen_scalable_2024,li_heterogeneous_2024},
strain tuning of the SnVs emission frequency 
\cite{clark_nanoelectromechanical_2024,brevoord_largerange_2025}, two-photon quantum interference of emitted photons \cite{arjonamartinez_photonic_2022,bushmakin_twophoton_2024} and their conversion to the telecom band
\cite{QFC}.
Moreover, multiple works have shown cavity-coupled SnVs with cooperativities up to 
$C \approx 15$
\cite{rugar_quantum_2021,%
kuruma_coupling_2021,%
herrmann_coherent_2024}.

In this work, we report on the fabrication of diamond photonic crystal cavities. 
We first characterize a subset of the fabricated devices in a room temperature setup (Appendix \ref{sec:room_temp_setup}), obtaining statistics on their quality factors and resonance frequencies. 
At cryogenic temperatures (Appendix \ref{sec:setup}), cavity-coupled SnV centers are identified through \textit{in-situ} frequency tuning using nitrogen gas deposition~\cite{mosor_scanning_2005, srinivasan_optical_2007}. 
We then focus on two devices, for which we measure both $C$ and $C_\s{coh}$.
The coherent cooperativity $C_\s{coh}$ quantifies the enhancement of the light-emitter interaction, including dephasing, and is a key figure of merit determining the fidelity of entanglement schemes \cite{borregaard_quantum_2019}.
So far, reaching $C_\s{coh} > 1$ for cavity-coupled SnVs has remained elusive.
Here, we measure a Purcell-reduced lifetime corresponding to \valueC~and an emitter-modulated cavity transmission, obtaining \valueCcoh.

\section{Design \& Fabrication}
A schematic illustration of our photonic crystal cavities is shown in Fig. \refsub{fig:1}{a}. The devices are single-sided photonic crystal cavities, consisting of a strong mirror, a central cavity region and a weak mirror \cite{nguyen2019integrated}.
For both strong and weak mirrors, we use the same lattice constant $a_n$ (Appendix \ref{sec:cavity-design}).
The total cavity decay rate $\kappa=\kappa_e+\kappa_i$ is divided into an external part $\kappa_e$ that is guided by the diamond waveguide and an internal part $\kappa_i$ that contains all other decay channels.
Fig \refsub{fig:1}{b} shows a finite element method (FEM) simulation of the electric field strength of the cavity mode. The highest cooperativity is achieved for a SnV center located at its maximum.
Fig. \refsub{fig:1}{c} illustrates a simplified level structure of the SnV–cavity system, with rates indicated by arrows and energies by dotted lines.

The realization of SnV center integrated devices is described in two main steps: first, the SnV centers are created via ion implantation and high temperature low pressure annealing; second, photonic crystal cavity devices are fabricated via the quasi-isotropic etch undercut method \cite{khanaliloo_high-qv_2015, mitchell2019realizing, mouradian_rectangular_2017, wan_large-scale_2020, rugar_quantum_2021, ruf_cavityenhanced_2021}.

Starting from a bulk diamond, the integration of SnV centers in diamond substrates follows the methods described in Appendix \ref{sec:fabrication-process}. We incorporate \ce{^{120}}Sn ions in the diamond substrate via high-energy ion implantation (350 keV, \ang{7} implantation angle), resulting in an implantation depth of $\SI{88}{\nano\meter}\pm\SI{16}{\nano\meter}$. 
This places the ions close to the vertical maximum of the cavity mode.
The SnV centers are activated via high-temperature, low-pressure annealing (\SI{1100}{\degreeCelsius} for a duration of 4 hours) by the surface-protected annealing method \cite{codreanu_diamond_2025}.

The device fabrication follows with patterning the \ce{Si_xN_y} hard mask material, followed by the transfer pattern into the diamond substrate and vertical coverage with \ce{AlO_x} of the structures sidewalls. Next, the devices are undercut employing quasi-isotropic etch (QIE), followed by an upward etch to thin the devices to a thickness of $\sim\SI{180}{\nano\meter}$. Fabrication concludes with an inorganic removal of the hard mask materials.
An exemplary fabricated photonic crystal cavity device is shown in Fig. \refsub{fig:1}{d}. Complementary scanning electron microscopy (SEM) and atomic force microscopy (AFM) inspection of the representative devices (Appendix \ref{sec:geometry-characterization}) allows us to infer the obtained device geometry, confirming a high degree of control over our fabrication process.
The cavities are grouped in arrays of six devices each, as shown in Fig. \refsub{fig:1}{e}. Within each cavity array, we sweep the number of holes in the weak mirror region (from one to six), to vary the cavity outcoupling $\kappa_{e}$. 
%Additionally, we demonstrate reproducibility of the proposed fabrication methods by comparing room temperature optical characterization statistics between devices on Sample 1 and Sample 2.

\section{Device Characterization}
\begin{figure*}[t]
    \centering
    \includegraphics[width=\linewidth]{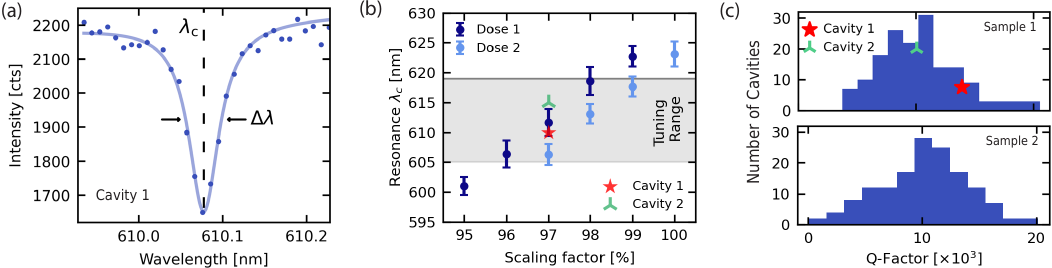}
    \caption{
    Room-temperature PCC characterization. 
    \textbf{(a)}
    The cavity resonance is measured by taking a reflection spectrum with a white supercontinuum laser.
    The spectrum is fitted with a Lorentzian plus linear background, from which the resonance wavelength $\lambda_c$ and full width at half maximum (FWHM) $\Delta\lambda$ are extracted to determine the quality factor $Q = \lambda_c / \Delta\lambda$.
    \textbf{(b)} Cavity resonance vs. device scaling factor for Sample 1.
    We vary the geometrical parameters of the PCCs by a common scaling factor, changing the hole diameter, lattice constant and device width. 
    % The scaling factor here shown is from 95 \% to 100 \% to ensure a blue-shifted resonance wavelength $\lambda_c$ distribution from $\sim\SI{600}{\nano\meter}$ to $\sim\SI{620}{\nano\meter}$, respectively.
    Devices within the indicated tuning range can be brought into resonance with the SnVs zero-phonon line (ZPL) at \SI{619}{\nano\meter} by gas deposition.
    (Appendix \ref{sec:gas-tuning}).
    We use two different e-beam exposure doses, \SI{265}{\micro\coulomb\per\centi\meter\squared} and 
    \SI{275}{\micro\coulomb\per\centi\meter\squared}, (dark and light blue, respectively).
    \textbf{(c)}
    Distribution of quality factors for Sample 1 (top panel) and Sample 2 (bottom panel).
    The average quality factor is \valueQavg.
    Due to the limited spectrometer resolution of
    \SI{0.021}{\nano\meter} the quality factors reported are a lower bound.
    } 
    \label{fig:2}
\end{figure*}
The optical characterization is performed first at room temperature by contact-coupling a tapered optical fiber and probing the devices' spectra in reflection with a supercontinuum white laser.
The reflection spectrum for a specific cavity, from here on called Cavity 1, is shown in \refsub{fig:2}{a}.
The data is fitted by a Lorentzian with linear background, and the quality factor is obtained as $Q = \lambda_c / \Delta\lambda$.
We show fabrication results on two distinct diamond samples (herein after Sample 1 and 2) and compare their quality factor statistics. For Sample 1, of the 232 investigated devices, 200 show a resonance. Similarly, in Sample 2, it is 216 out of 254 devices.
This corresponds to a fabrication yield of $\sim 85\%$ for both samples.
We report on \valueNumQ{} quality factors in total (Appendix \ref{sec:device-stats}).

To compensate cavity resonance shifts from fabrication process variability and yield cavity devices with resonances $\lesssim\SI{619}{\nano\meter}$, we sweep the cavity design between neighboring cavity arrays by varying the scaling factor (lattice constant $a_n$, waveguide width $w_\s{wg}$ and hole radius $r$).
Fig. \refsub{fig:2}{b} shows the cavity resonances as a function of this scaling factor for Sample 1.
The nominal design parameters (corresponding to scaling factor 100\%) show a cavity resonance slightly exceeding \SI{620}{\nano\meter}, in good agreement with FEM simulations (Appendix \ref{sec:cavity-simulations}). 
The scaling factor is varied around the nominal 100\% such that a distribution of cavity resonances around the SnV zero-phonon line (ZPL) is obtained. Specifically, devices with a scaling factor from 95\% to 99\% show a cavity resonance blue-detuned with respect to the SnV ZPL, allowing for gas tuning of the cavity resonance (Appendix \ref{sec:gas-tuning}).

Additionally, for Sample 1, we use two electron beam (e-beam) exposure doses. We find that devices exposed with a higher e-beam exposure dose
(\SI{275}{\micro\coulomb\per\centi\meter\squared}, light blue) show on average a $\sim\SI{5}{\nano\meter}$ lower cavity resonance than the devices exposed with a lower e-beam exposure dose (\SI{265}{\micro\coulomb\per\centi\meter\squared}, dark blue), all other fabrication process parameters are kept equal. This shows a high sensitivity of the cavity resonance to a small variation of the process parameters, demonstrating that a high degree of control over the fabrication process is required to ensure high accuracy over the target cavity resonance.
Two cavities are marked and investigated in more detail in the later sections of this work.

Fig. \refsub{fig:2}{c} shows the distribution of quality factors for Sample 1 (top) and Sample 2 (bottom). The average quality factor for both samples is \valueQavg. No significant dependence is observed between the measured device quality factors and the respective scaling factors. Both samples show a similar distribution of quality factors, demonstrating the reproducibility of the proposed fabrication methods.
The spectrometer resolution is \SI{0.021}{\nano\meter} corresponding to a maximally measurable quality factor of 
$Q \approx 3 \times 10^4$.
Due to this limit, the measured quality factors are a lower bound.
In the next section, the quality factors for Cavities 1 and 2 are more accurately determined through resonant laser scans.

These results show that the developed fabrication process is characterized by a high control over the process parameters and systematically yields high-performance cavity devices.

\section{Cavity-Emitter Coupling}
\begin{figure*}[t]
    \centering
    \includegraphics[width=\linewidth]{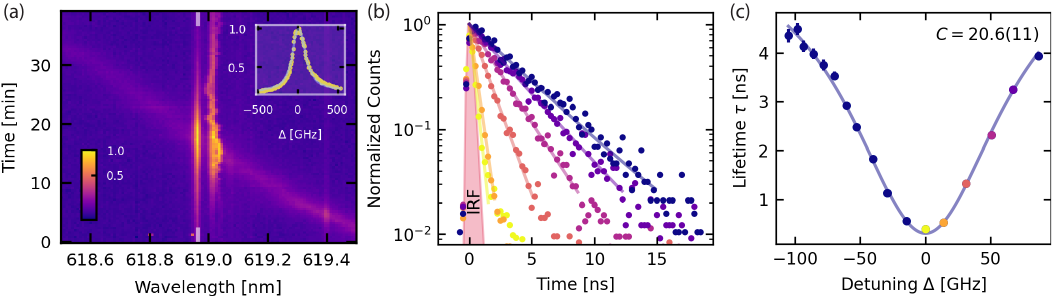}
    \caption{
    Cavity-emitter coupling. 
    \textbf{(a)}
    Detected counts on a spectrometer as the cavity is tuned. This experiment is used to identify cavity-coupled SnVs.
    Initially, the cavity resonance is shifted to 
    $\sim \! \SI{619.5}{\nano\meter}$ using gas deposition.
    Then, the cavity is tuned backwards by evaporating the deposited nitrogen with a green laser.
    When the cavity is on-resonance with the SnV (Emitter 2 indicated by the gray marks), it increases in brightness.
    The inset shows the spectrometer counts at this frequency ($\SI{618.96}{\nano\meter}$) as a function of cavity detuning $\Delta$.
    \textbf{(b)}
    The SnV is excited off-resonantly with a pulsed green laser ($\SI{70}{\pico\second}$ pulse length).
    The collected ZPL counts are recorded on a timetagger and the result is fitted with a single exponential decay to obtain the lifetime $\tau$.
    \textbf{(c)}
    Lifetime as a function of detuning fitted with $\tau = \tau_0 / (C \, f(\Delta) + 1)$, leading to \valueC.
    }
    \label{fig:3}
\end{figure*}
%Might need one sentence to introduce the section
The Purcell reduced lifetime of an emitter in an optical cavity is given by
\begin{equation}
    \tau = \frac{\tau_0}{C \, f(\Delta) + 1},
\end{equation}
where $f(\Delta) = 1 / (1 + 4 \Delta ^2/\kappa^2)$ is the spectral mismatch factor, $\Delta \coloneqq \omega_c - \omega_a$ is the detuning between cavity and SnV, and $\tau_0$ is the natural lifetime.
The cooperativity $C$ follows
\vspace{-3mm}
\begin{equation}
    C = F_p \, \beta_0 \, u^2(\mathbf r) \, \zeta,     
\end{equation}

where $F_p=\frac{3}{4\pi^2}\left(\frac{\lambda}{n}\right)^3\frac{Q}{V}$ is the Purcell factor, with $V$ the mode volume, $\beta_0\approx0.36$ \cite{herrmann_coherent_2024, gorlitz_spectroscopic_2020, iwasaki_tinvacancy_2017, rugar_quantum_2021} is the probability for the SnV to spontaneously decay via the C transition of the ZPL,
$u(\mathbf r) \in [0,1]$ 
is the normalized field strength at the emitter location, and
$\zeta$ is the polarization mismatch.
The SnV transition dipole is oriented along the diamond's $\langle 111 \rangle$ crystallographic axis, whereas the cavities are oriented along one of the $\langle 110 \rangle$ axes. At the field's maximum, $\zeta=2/3$ (Appendix \ref{sec:cavity-simulations}).
The value of $u(\mathbf r)$ is randomly distributed, due to the blanket implantation used in this case.

To find an SnV that is coupled to the cavity we perform the measurement shown in Fig. \refsub{fig:3}{a}. First, the sample is cooled down in a cryostat to \SI{9}{\kelvin}. We then
introduce a small amount of nitrogen gas into the vacuum chamber
(Appendix \ref{sec:gas-tuning}).
The gas deposits on the PCC and shifts the cavity resonance 
$\lambda_\s{c}$.
We stop the process slightly red-detuned from the SnV's ZPL.
Then we take spectra alternating between excitation with the white laser and the green laser, both of which are sent from the top with an objective. 
A fraction of the emission of the SnV and of the light scattered into the cavity is guided by the waveguide and collected with a tapered fiber, see Fig. \refsub{fig:1}{a}.
With the white laser, we measure $\lambda_\s{c}$ and the green laser
(\SI{4.5}{\milli\watt}, \SI{10}{\second} per step)
backtunes the cavity and off-resonantly excites the SnVs.
A cavity-coupled SnV is detected as an increase in counts while the cavity is in resonance with the SnV.
This measurement was performed on 9 cavities in total, out of which 7 show cavity-coupled emitters. The data shown in Fig. \refsub{fig:3}{a} corresponds to Cavity 2 and the line with gray marks corresponds to Emitter 2.
All other data of the main text belongs to Emitter 1 in Cavity 1.

After identifying a cavity-coupled SnV, we measure its Purcell reduced lifetime $\tau$ to obtain the cooperativity $C$.
In Fig. \refsub{fig:3}{b}, we again tune the cavity, this time using less green laser power 
(\SI{2.7}{\milli\watt}, \SI{1}{\second} each step)
to achieve smaller frequency steps.
At each step, the SnV is excited with a 
pulsed green laser ($70$~ps pulse length) and the ZPL emission is recorded on a time tagger. 
We spectrally filter the collected signal to isolate the ZPL of the SnV under investigation, see Appendix \ref{sec:setup}.
The measured counts are fitted with a single exponential decay to obtain the lifetime.
In this case, the smallest value is
$\tau_\mathrm{min} = \SI{0.39 \pm 0.03}{\nano\second}$.
This is on the order of the instrument response function (IRF) and is limited by the approximately
\SI{400}{\pico\second} jitter of our avalanche photodiode.

Fig. \refsub{fig:3}{c} shows all measured lifetimes as a function of the cavity-emitter detuning
$\Delta = \omega_c - \omega_a$. 
The data is fitted with 
$\tau = \tau_0 / (C \, f(\Delta) + 1)$ obtaining 
$\tau_0 = \SI{6.5 \pm 0.2}{\nano\second}$
and
\valueC.
Here, the lowest point is excluded from the fit due to the IRF limit.

These results show a cavity-coupled SnV device, with a strongly Purcell-reduced lifetime.
However, lifetime measurements are insensitive to dephasing effects.
To further probe the coherence of this light-matter interface, we perform resonant measurements of the same SnV in the next section.

\section{Coherent Interactions}
\begin{figure*}[t]
    \centering
    \includegraphics[width=\linewidth]{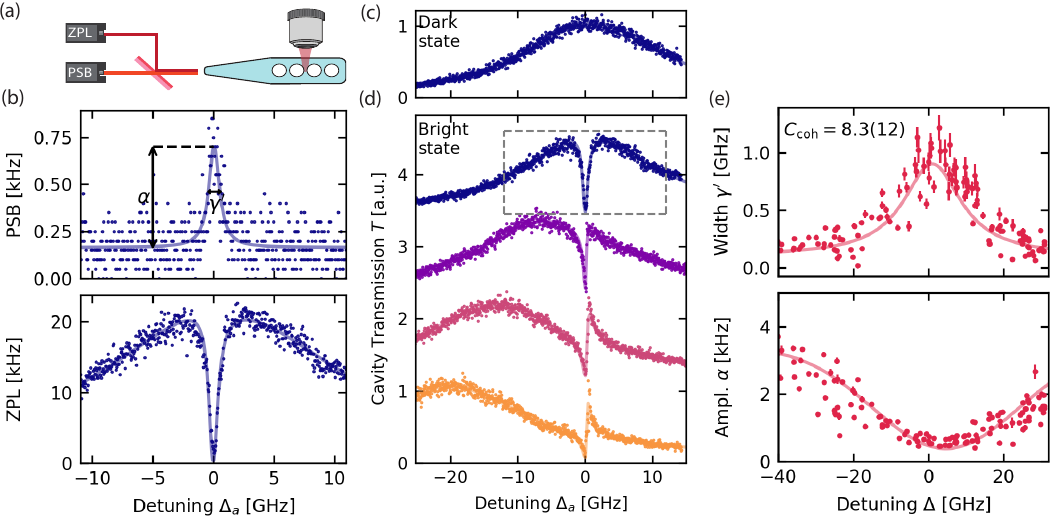}
    \caption{
    Coherent SnV-cavity interactions.
    \textbf{(a)}
    The collected light is split into ZPL and PSB using a dichroic mirror.
    \textbf{(b)}
    The cavity and SnV are tuned on resonance where $\Delta \equiv \omega_a - \omega_c = 0$.
    The red laser is scanned across the cavity resonance, PSB and ZPL are recorded as a function of the laser-SnV detuning $\Delta_a \equiv \omega - \omega_a$.
    The top panel shows the collected PSB counts:
    A Purcell broadened lineshape that is fitted by a Lorentzian with amplitude $\alpha$ and linewidth $\gamma^\prime$.
    The bottom panel shows the emitter-modulated cavity transmission \cite{reiserer_cavitybased_2015}.
    For this detuning, Equation \eqref{eq:transmission} predicts a symmetric dip. 
    Fitting the data reveals a contrast of $\Delta T =\valueDeltaT$.
    \textbf{(c)}
    When the SnV is in a dark state, we measure the Lorentzian transmission of a bare cavity, where \valueQ.
    \textbf{(d)}
    When the SnV is in the bright state, the cavity transmission is modulated by the interaction with the emitter.
    The transmission is shown for four different detunings 
    $\Delta \in [0, 5, 12, 20] \, \si{\giga\hertz}$.
    \textbf{(e)}
    The top panel shows the emitter linewidth $\gamma^\prime$ as a function of detuning.
    Due to the Purcell effect, the linewidth broadens as the emitter and cavity are tuned on-resonance.
    The data is fitted with $\gamma^\prime = \gamma (f(\Delta) \, C_\s{coh} + 1)$, and yields \valueCcoh.
    Conversely, the bottom panel shows that the amplitude $\alpha$ reduces.
    This can be explained by a lower excited state population on resonance
    (Appendix \ref{sec:cqed_modelling}).
    }
    \label{fig:4}
\end{figure*}
Dephasing is an important factor in predicting the performance of a cavity-emitter system for entanglement protocols
\cite{borregaard_longdistance_2015}.
While the cooperativity $C=\frac{4g^2}{\kappa\gamma_0}$ quantifies the cavity-enhanced decay rate of the emitter over its natural linewidth $\gamma_0$, the coherent cooperativity $C_\s{coh}=\frac{4g^2}{\kappa\gamma}$ quantifies the cavity-enhanced emitter linewidth over its off-resonant linewidth $\gamma$.
Here, $g$ is the emitter-cavity coupling and $\gamma=\gamma_0+\gamma_\s{dep}$ is the off-resonant emitter linewidth including dephasing.
% The lifetime of an emitter is not impacted by dephasing.
% Instead, to probe dephasing effects, we perform resonant transmission experiments.

In Fig. \ref{fig:4}, we probe coherent cavity-emitter interactions by scanning a red laser across the cavity resonance.
The light collected from the cavity is split into zero-phonon line (ZPL) and phonon side band (PSB) using a dichroic mirror (Fig. \refsub{fig:4}{a}).
We simultaneously measure the Purcell-broadened linewidth
$\gamma^\prime$ of the emitter in the PSB and the modulated cavity transmission $T(\omega)$ in the ZPL.
It is important to note that the PSB signal is relatively weak because a significant amount of the PSB is reflected off the PCC outcoupling mirror
(Appendix \ref{sec:cavity-simulations}).
Solving the Hamiltonian of the system in the weak excitation regime (Appendix \ref{sec:cqed_modelling}) gives an analytical expression for the transmission
\begin{equation}
\label{eq:transmission}
    T(\omega) = 
    \left|
    \frac{
    \sqrt{\kappa_\s{in}\kappa_e}
    }{
        \kappa/2 
        + i (\omega - \omega_\s c) 
        + \frac{g^2}{\gamma/2 + i(\omega - \omega_\s a)}
    }
    \right|^2,
\end{equation}

where $\omega_c$, $\omega_a$ and $\omega$
are the frequencies of the cavity, atom and probe laser, respectively
\cite{reiserer_cavitybased_2015}.
Fig. \refsub{fig:4}{b} shows the emitter lineshape (top) and cavity transmission (bottom) when emitter and cavity are on resonance ($\Delta = 0$).
We find a transmission dip contrast of
$\Delta T = \valueDeltaT$, a significant improvement over the 0.25 previously measured with a waveguide-coupled SnV
\cite{pasini_nonlinear_2024}.
\begin{figure*}[t]
    \centering
    \includegraphics[width=\linewidth]{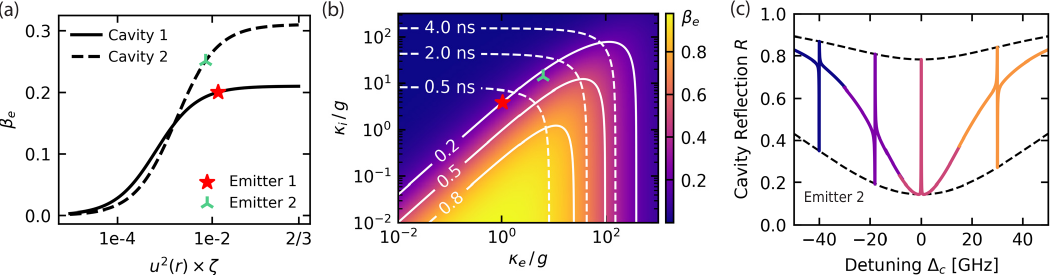}
    \caption{
    Influence of the external coupling.
    \textbf{(a)}
    The external beta factor $\beta_e = \frac{\kappa_e}{\kappa}\frac{C}{C+1}$ as a function of the emitter location for two different external couplings $\kappa_e/\kappa \in \{0.21, 0.31\}$, corresponding to Cavity 1 and Cavity 2.
    Within each curve, we mark the point corresponding to the investigated emitters.
    \textbf{(b)}
    The external beta factor
    as a function of the loss rate into the waveguide $\kappa_e$ and other losses $\kappa_i$.
    The dashed lines mark a Purcell reduced lifetime of 0.5, 2 and 4 \si{\nano\second} to a natural lifetime of $\tau_0 = \SI{6.5}{\nano\second}$.
    The solid lines an external beta factor of $0.2, 0.5$ and $0.8$.
    A similar plot for the silicon vacancy center (SiV) can be found in 
    \cite{knall_efficient_2022}.
    \textbf{(c)}
    Calculated cavity reflection as a function of laser-cavity detuning $\Delta_c$ for four different emitter frequencies.
    The broad dip corresponds to the response of the cavity and the peak-dip pairs correspond to the emitters' response, similar to those found in Fig. \refsub{fig:4}{d}, but in reflection.
    The dashed black lines mark the envelope of the peaks and the dips.
    We use the C-QED parameters of Emitter 2. 
    }
    \label{fig:5}
\end{figure*}

Before each laser scan, we apply a 
\SI{100}{\micro\second} green laser pulse.
This pulse probabilistically initializes the SnV to the bright SnV$^-$ state.
Conversely, the green pulse can also ionize the SnV$^-$ into a dark state
\cite{gorlitz_coherence_2022}.
We use the PSB signal to discern whether the SnV is in a dark or bright state.
When the emitter is in a dark state (Fig \refsub{fig:4}{c}), we observe the Lorentzian transmission of a bare cavity.
Fitting this data gives a value for the quality factor of \valueQ{}.
This value is more accurate compared to Fig. \ref{fig:2}, as it is not limited by the spectrometer resolution.
When the emitter is in the bright state, the cavity transmission is modulated by the emitter.
Fig \refsub{fig:4}{d}, shows the cavity transmission for four different detunings $\Delta \in [0, 5, 12, 20] \, \si{\giga\hertz}$, where the $\Delta=0$ case corresponds to the data in Fig \refsub{fig:4}{b} (dashed rectangle).

In Fig. \refsub{fig:4}{e} the linewidths $\gamma^\prime$ of the PSB signal are plotted as a function of detuning and fitted with
$\gamma^\prime = \gamma (f(\Delta) \, C_\s{coh} + 1)$. 
This reveals cavity quantum electrodynamics (C-QED) parameters $\{g, \kappa, \gamma\}/2\pi = \{\SI{1.94\pm 0.08},\, \SI{19\pm 0.3}, \,\SI{0.0975\pm 0.0004}\}$~GHz yielding a coherent cooperativity of
\valueCcoh.
The emitter presents additional broadening compared to the natural linewidth of 
$\gamma_0=2\pi\cdot\SI{24.5 \pm 0.8}{\mega\hertz}$
likely due to a combination of dephasing and temperature broadening.
At the temperature of the experiment (\SI{9}{\kelvin}), a linewidth two to three times broader than the lifetime limit is expected \cite{wang_transformlimited_2024}.
Reduced temperature broadening alone would roughly double $C_\s{coh}$ if the experiment were performed at \SI{4}{\kelvin}.
The amplitude $\alpha$ of the PSB signal decreases when the emitter and the cavity are on resonance.
This is a result of a decrease in the steady-state population of the emitter's excited state due to an increased decay rate into the cavity at resonance.
We fit the data with an envelope of the scattering function of the emitter (Appendix \ref{sec:cqed_modelling}). For the fit, we fix $g, \kappa$ and $\gamma_0$ to the measured values and vary the proportionality constant and the center of the curve.

We attempted to measure the coherent cooperativity for a total of four SnVs in four different cavities.
The emitters were investigated because they increase in brightness when in resonance with the cavity, as shown in Fig. \refsub{fig:3}{a}.
Of the four emitters, one showed no transmission dip,
one showed a slight dip with $C_\s{coh} \approx 0.2$
and the last two showed significant dips (Cavities 1 and 2).
The large variance in $C_\s{coh}$ is likely due to the random position of the SnV in the cavity mode, quantified by $u(\mathbf r)$.
The candidate that showed no transmission dip at all might suffer from excessive spectral diffusion, such that the dip washes out over the timescale of the laser scan (\SI{20}{\second}).
Together with a laserscan in reflection (Appendix \ref{sec:reflection}) to measure the ratio $\kappa_e/\kappa$, this section completes a full investigation of C-QED parameters
(Appendix \ref{sec:cqed_parameters}).
All measurements of the main text are repeated for Cavity 2 
(Appendix \ref{sec:device-two}).

\section{Performance Characteristics}
An important performance characteristic for cavity-enhanced light-matter interfaces is the external beta factor $\beta_e = \frac{\kappa_e}{\kappa}\frac{C}{C+1}$.
It is the probability that an excited emitter will emit into the cavity mode and that the emitted photon will leave the cavity through the outcoupling mirror.
In Fig. \refsub{fig:5}{a} we plot $\beta_e$ as a function of the factor $u^2(\mathbf r)\times \zeta$ (see Appendix \ref{sec:cavity-simulations}), which effectively links the emitter's location with the cavity-emitter coupling $g$ for a fixed cavity design.
We do this for two external couplings $\kappa_e/\kappa \in \{0.21, 0.31\}$ that correspond to Cavity 1 and Cavity 2, respectively.
For each curve, we mark the investigated emitters.
For these emitters, further improving the emitter location would not have a large effect on $\beta_e$ because $C/(C+1)$ is already close to 1.
However, increasing the outcoupling $\kappa_e/\kappa$ would also increase $\beta_e$ in the limit of well-positioned emitters.

In Fig. \refsub{fig:5}{b}, we illustrate how $\beta_e$ 
depends on the internal and external cavity losses.
The external loss $\kappa_e$ is a parameter that can be chosen by design:
The fewer holes in the weak mirror region, the larger $\kappa_e$.
Increasing $\kappa_e$ will increase the ratio of photons leaving the cavity through the weak mirror.
Simultaneously, it will decrease the quality factor $Q$ and thus the probability of emitting into the cavity mode
($C \propto F_p \propto Q \propto 1/\kappa $).
In effect, there is an optimum $\kappa_e$ for a fixed $\kappa_i$ \cite{knall_efficient_2022}.
The internal loss $\kappa_i$ can be seen as a gauge for how well the cavity is designed and fabricated, and decreasing $\kappa_i$ will always result in a higher $\beta_e$.
In Fig. \refsub{fig:5}{b} we also show iso-lifetimes curves for an emitter with $\tau_0 = \SI{6.5}{\nano\second}$.
Short lifetimes increase the complexity of emission-based experiments due to a trade-off between the need for short excitation pulses and their ability to selectivly address the optical transition of each spin state.

The external coupling $\kappa_e/\kappa$ is also an important parameter for entanglement protocols that do not rely on emission, but on reflection from the cavity.
In Fig. \refsub{fig:5}{c} we show the calculated reflection for Cavity 2 for different emitter frequencies. 
We use the C-QED parameters of Emitter 2.
For recently proposed protocols \cite{knaut_entanglement_2024}, the entanglement fidelity is directly dependent on how small the reflection at the emitter dip is.
It can be mathematically proven that there always exists a cavity-emitter detuning for which the dip goes to zero as long as the cavity is overcoupled, i.e., $\kappa_e/\kappa>0.5$.
This is illustrated in Fig. \refsub{fig:5}{c}, where we see that it is not possible to reach a dip to zero with our maximum outcoupling of $0.31$.

\section{Outlook}
The presented cavities reach coherent cooperativities compatible with quantum networking protocols \cite{knaut_entanglement_2024, ruskuc_multiplexed_2025}.
However, several opportunities remain to further enhance device performance.
Increasing the outcoupling $\kappa_e/\kappa$ would benefit both emission-based and reflection-based entanglement protocols.
Specifically, a cavity with the same internal loss as presented here could reach external beta factors as high as $\beta_e=0.67$ when $\kappa_e$ is chosen optimally.
Increasing $\kappa_e$ can be done by further removing unit cells or with alternative cavity designs \cite{knall_efficient_2022,bopp_sawfish_2024}. 
Alternatively, fabricating cavities with smaller internal losses $\kappa_i$ makes it possible to achieve overcoupled cavities without compromising their quality factor $Q$ \cite{ding_highq_2024}.
Emerging diamond fabrication processes \cite{guo_directbonded_2024} result in ultra-low surface roughness of the device layer \cite{guo_tunable_2021} and have demonstrated the highest optical quality factors for one- and two-dimensional photonic crystal cavities to date \cite{ding_highq_2024, ding_purcellenhanced_2025}.

Using adiabatic fiber coupling instead of lensed fiber coupling could increase the photon collection roughly threefold \cite{pasini_nonlinear_2024, zeng_cryogenic_2023}.
To successfully implement this, the device containing the SnV center must be mechanically isolated from the coupling fiber and the fiber must be fully thermalized to the sample temperature.
Finally, methods such as focused ion beam (FIB) implantation \cite{schroder2017scalable} and masked implantation \cite{burek_freestanding_2012} can be implemented to gain more control over the position of the SnV within the cavity mode.
These prevent the formation of SnVs in all regions of the device (taper, waveguide, and low-field-strength regions), with the exception of the target position in the region of maximum electric field in the cavity mode.
Moreover, FIB implantation with sufficient accuracy and precision \cite{Cheng2025-ve} enables positioning the SnVs close to the mode maximum, thereby yielding higher and more consistent cooperativities.

Combining the above improvements gives access to a parameter regime with near unity external coupling factors $\beta_e$ while retaining spin-selective excitation.
Such a device would enable various applications, such as high-rate entanglement distribution or even the deterministic creation of multi-dimensional cluster states
\cite{borregaard_oneway_2020}.

\begin{acknowledgments}
    We thank Julius Fischer, Yanik Herrmann and Niv P. Bharos for helpful and important discussions, Henri Ervasti for software support, Hans K. C. Beukers and Christopher Waas for help with measurement scripts, Julia M. Brevoord and Alexander M. Stramma for help with the cryogenic setup, and Julius Fischer and Yanik Herrmann for proofreading the manuscript.
    We acknowledge financial support from the joint research program “Modular quantum computers” by Fujitsu Limited and Delft University of Technology co-funded by the Netherlands Enterprise Agency under project number PPS2007, from the Dutch Research Council (NWO) through the Spinoza prize 2019 (project number SPI 63-264), from the Dutch Ministry of Economic Affairs and Climate Policy (EZK) as part of the Quantum Delta NL programme, from the Quantum Internet Alliance through the Horizon Europe program (grant agreement No. 101080128) and from The Kavli Foundation through the Kavli Institute Innovation Award "Quantum Materials for Broad-Band Quantum Transduction".

    \textbf{Author contributions:} N.C., T.T and D.B.R. contributed equally to this work. N.C., M.R. and L.D.S. designed the devices. N.C. developed and optimized the device fabrication process with assistance from M.R. and S.G. N.C. fabricated the devices and tapered fibers, with assistance from L.G.C.W. and C.E.S. N.C. and M.P. designed the room temperature setup and N.C. and L.D.S. built it. N.C. conducted the room temperature experiments and N.C. analyzed the data with help from T.T. M.P. and T.T. built the cryogenic setup. T.T. built and tested the gas tuning system. T.T. and D.B.R. conducted the cryogenic experiments and analyzed the data. D.B.R. and T.T. modelled the fabricated devices. C.F.P. characterized the diamond samples before fabrication. N.C., T.T., D.B.R. and R.H wrote the manuscript with input from all authors. R.H. supervised the experiments.

    \textbf{Data availability:} The data that support the findings of this article are openly available \cite{codreanu_data_2025}.
    
\end{acknowledgments}
\include{appendix}

\bibliography{references, references2}

\end{document}

%% file: appendix.tex
\appendix
\section{Room-Temperature Setup}
\label{sec:room_temp_setup}
\begin{figure}[ht]
\centering
\includegraphics[width=\linewidth]{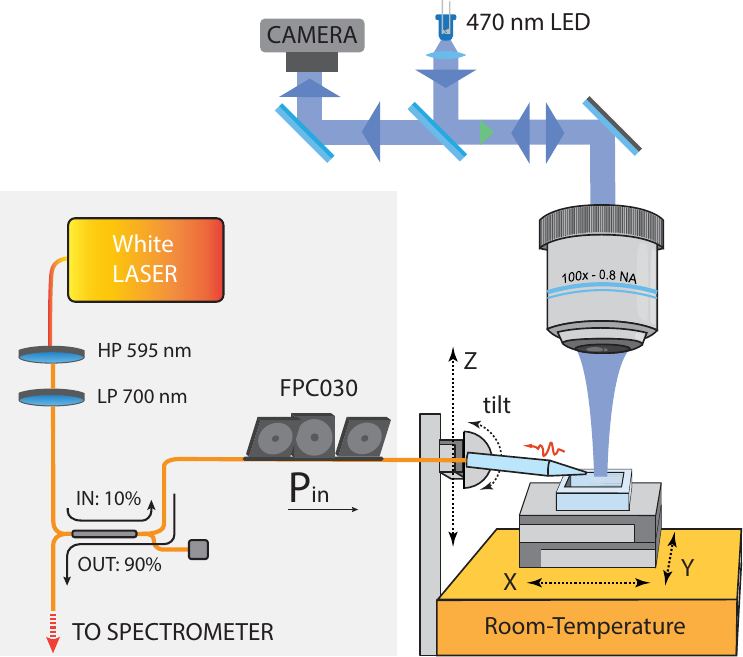}
\caption{Room temperature characterization setup}
\label{PCC_fig_RTNano}
\end{figure}
A schematic of the room temperature single-sided cavity devices characterization setup is shown in Figure \ref{PCC_fig_RTNano}. 

The diamond sample is mounted (via carbon tape) on the sample stage, positioned on top of the XY piezo positioner stack (Smaract MCS2-S), allowing sample position control in the XY direction. The tapered optical fiber is mounted on a variable angle (manually set) stage, mounted on a slip-stick Z piezo positioner (Smaract MCS2-S) which allows one to control the vertical position of fiber. %This allows for fast characterization by broad-band probing reflection probing of single-sided resonant devices. 

The sample is imaged by illuminating it with a LED at $\sim\SI{470}{nm}$ using a microscope objective (100X, numerical aperture (NA) 0.8), coupled to a digital camera (CS165CU/M, Thorlabs). This provides enough resolution to optically resolve both the optical fiber tip and the free-hanging cavity devices coupling tapered waveguide.

The input broad-band laser light source (SuperK Compact, NKT Photonics supercontinuum white light source) is delivered adiabatically via the contact coupled tapered optical fiber. This is filtered by cascading a Long Pass and a Short Pass filters, such that the input probing light is comprised between \SI{595}{\nano\meter} and \SI{700}{\nano\meter}, within the cavity resonance spectral region of interest. A 90:10 beam splitter is used to send $10\%$ of the laser light to the input fiber, where we use a Fiber Polarization Controller (FPC030 by Thorlabs) to vary the polarization and match it to the device polarization (set by geometry).
The light couples to the single-sided cavity device via the weak mirror region and is reflected back to the same fiber port. The reflected signal (90\%) is then sent to the spectrometer detection (SpectraPro HRS-500).

\section{Cryogenic Setup}
\label{sec:setup}
\begin{figure*}[ht]
    \centering
    \includegraphics[width=\linewidth]{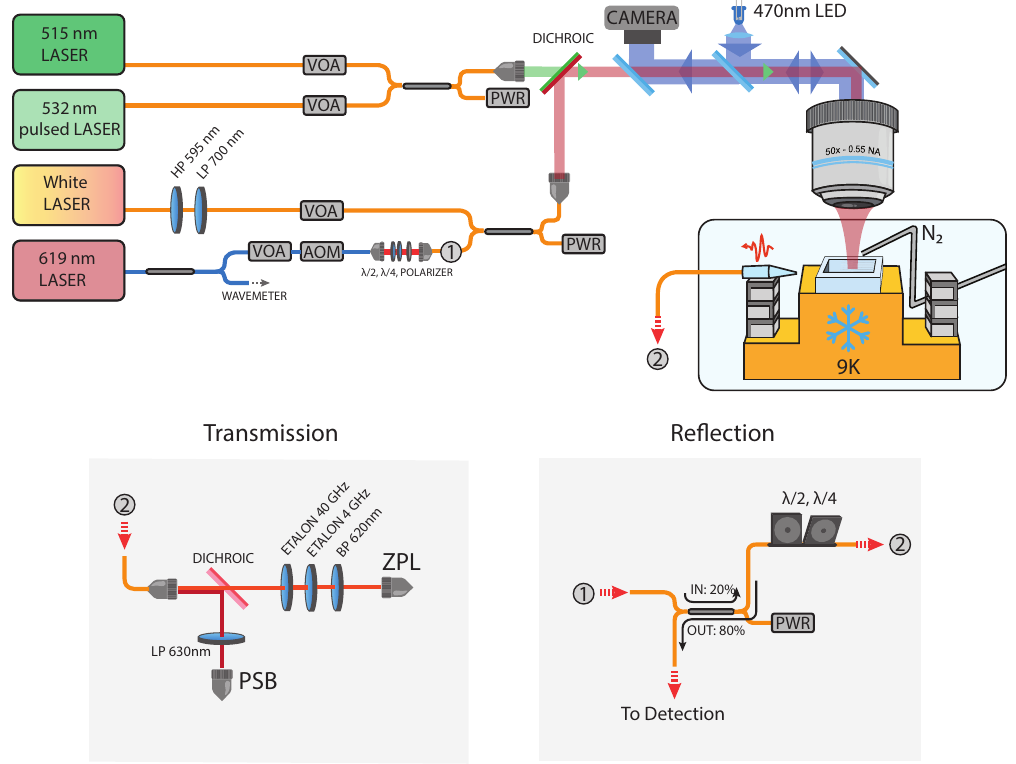}
    \caption{
    The measurement setup.
    For all experiments in the main text, the PCCs are measured in transmission, where the tapered fiber is directly connected to the detection setup shown in the bottom left.
    For the reflection measurements (Supplement Section \ref{sec:reflection} the setup show in the bottom right is plugged in between the red laser and the detection setup.
    }
    \label{fig:setup}
\end{figure*}

Fig. \ref{fig:setup} shows a sketch of the measurement setup.
The diamond sample is cooled to $\sim \SI{9}{\kelvin}$ in a closed-cycle cryostat (Montana Instruments s50).
The usual \SI{4}{\kelvin} base temperature of the s50 is not reached because the $\s N_2$ gas line (Quantum Design, MI-4100-1184) is inside the heatshield and not thermalized to the \SI{40}{\kelvin} stage.
This can be improved straightforwardly if temperature is a concern in future experiments.
The sample is imaged through an objective (Mitutoyo 50X Plan Apochromat, NA: 0.55, WD: 13 mm).
The objective is outside the cryostat, the image is taken through a window in the top of the cryostat and recorded on a camera (Thorlabs, Zelux).
Besides the imaging path, all probe lasers are sent through the objective.
The green cw laser (Cobolt MLD-06 515 nm) and green pulsed laser (NKT Onefive Katana, 532 nm, 65ps) are combined with a 50/50 fiber beamsplitter.
One of the outputs is used for power calibration and the other one is sent to the sample.
Similarly, the red cw-laser (Toptica TA-SHG Pro, 619 nm) and white laser (NKT SuperK compact) are also combined on a 50/50 fiber beamsplitter.
The white laser is spectrally filtered using a 600 nm longpass and 700 nm shortpass filter.
Collection is performed using a tapered optical fiber (Thorlabs, SM600).
The fiber is tapered in $40\%$ hydrofluoric acid \cite{burek_fibercoupled_2017}.
For the transmission measurements (all measurements in the main text), the tapered fiber is passed through a vacuum-tight Teflon feedthrough and connected to the detection setup (Fig. \refsub{fig:setup}{b}).
Here, the collected light is split into PSB and ZPL using a dichroic beamsplitter (Semrock, FF625-SDi01).
The PSB light is filtered by a tunable long-pass (Semrock, TLP01-628).
For the lifetime measurements (Fig. \ref{fig:3}), we use off-resonant pulsed excitation.
That means we can excite multiple SnVs within the objective focus.
The ZPL path needs to spectrally filter the collected light to contain only the emission of the SnV we want to probe.
Therefore, we use a combination of a 40 GHz etalon (LightMachinery), a 4 GHz etalon (LightMachinery, OP-7423-843-1) and a 620 nm bandpass (Thorlabs, FBH620-10).
The 4 GHz etalon can drift on the timescale of tens of minutes.
We have placed it on a rotation stage (Thorlabs, K10CR1/M) and continuously reoptimize the angle.
For the reflection measurements (Supplement Section \ref{sec:reflection}), we probe the PCC with the red laser from the tapered fiber.
During these measurements, we insert the setup shown in Fig. \refsub{fig:setup}{c} between the red laser and the tapered fiber.
A fiber beamsplitter sends 20\% of the probe light to the sample and collects 80\% of the emission.
The polarization of the probe light is controlled with a motorized fiber paddle (Thorlabs, MPC320) 
The collected light is sent to the usual detection setup.

\section{Cavity Design and Device Layout}
\label{sec:cavity-design}
\begin{figure*}[ht]
    \centering
    \includegraphics[width=0.90\textwidth]{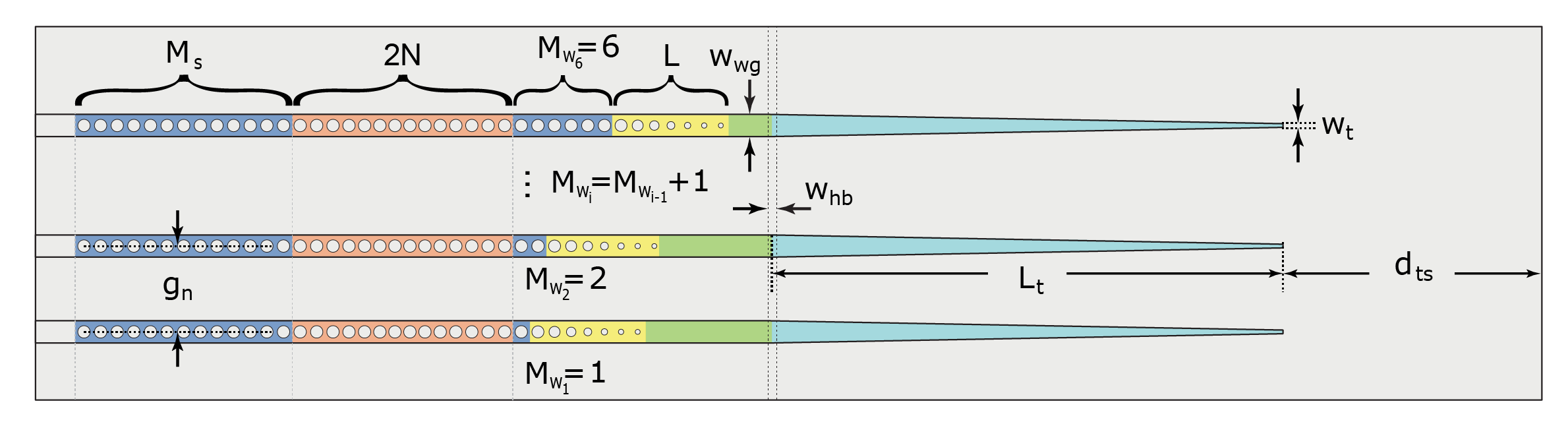}
    \caption{Schematic illustration of free-hanging single-sided photonic crystal cavity design (inspired from Ref. \cite{nguyen2019integrated}), in an array chiplet configuration.}
    \label{PCC_fig1}
\end{figure*}

\begin{figure}[ht]
    \centering
    \includegraphics[width=\linewidth]{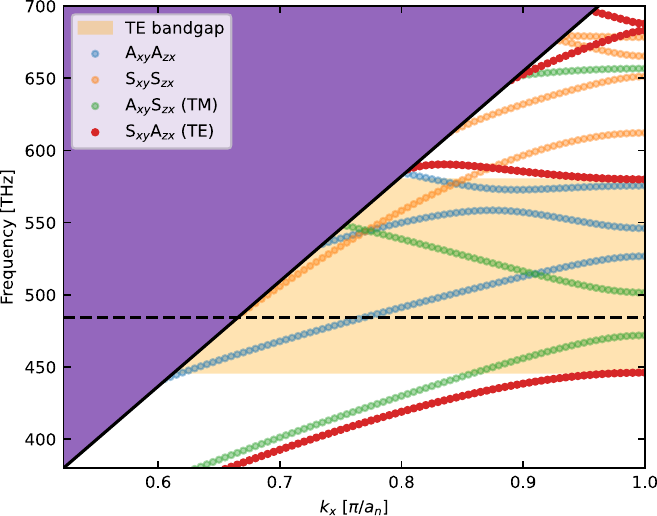}
    \caption{
    Bandstructure of the designed cavity. We classify the modes depending on the symmetry of the tangential electric field with respect to two planes. The dashed line representes the SnV's ZPL emission frequency. Axes defined in Fig. \refsub{fig:cavity-sim}{b}. Alternative notation for the mode symmetry is $\s{E}_{(o, n)} \leftrightarrow \s{S}_\s{xy}\s{A}_\s{zx}$, $\s{E}_{(e, n)} \leftrightarrow \s{S}_\s{xy}\s{S}_\s{zx}$, $\s{M}_{(e, n)} \leftrightarrow \s{A}_\s{xy}\s{S}_\s{zx}$ and $\s{M}_{(o, n)} \leftrightarrow \s{A}_\s{xy}\s{A}_\s{zx}$, with $n=1,2,\ldots$ the order of the mode \cite{joannopoulos_photonic_2011}.
    }
    \label{fig:cavity-design}
\end{figure}

The design of the symmetric photonic crystal cavity version was developed in 
\cite{ruf_cavityenhanced_2021}. The geometrical parameters are summarized in Table \ref{tab:cav_geom_params}.
\begin{table}[h!]
    \caption{Cavity design parameters}
    \label{tab:cav_geom_params}
    \centering
    \begin{tabular}{|c|c|c|}
        \hline
        Parameter & Symbol & Value\\
        \hline
        \hline
        Lattice constant & $a_n$ & $206$~nm\\
        \hline
        Hole radius & $r$ & $71$~nm \\
        \hline
        Width & $w_\s{wg}$ & $275$~nm \\
        \hline
        Thickness & $t_h$ & $230$~nm\\
        \hline
        Fractional deviation & $d$ & $0.1$\\
        \hline
        Slope well function & $\eta$ & $1.05$\\
        \hline
        Unit cells defect region & $N$ & $7$\\
        \hline
        Unit cells strong mirror & $M\text{s}$ & $13$\\
        \hline
        Unit cells weak mirror & $M_\text{w}$ & $1-6$\\
        \hline
    \end{tabular}
\end{table}
The defect is introduced with a functional form introduced in \cite{chan2012laser}, where the lattice constant is adiabatically varied on one side of the defect region following
\begin{equation}
    a_i = a\left(1-d(2y_i^3-3y_i^2+1)\right), \qquad i=0,..., N,
\end{equation}
where $y_i$ is given by
\begin{equation}
y_i=
    \begin{cases} 
      \frac{1}{2}(2x_i)^\eta, & x_i\leq 0.5, \\
      1-\frac{1}{2}\left(2(1-x_i)\right)^\eta, & x_i>0.5,
      \label{eq:map}
   \end{cases}
\end{equation}
with $x_i=i/N$. The other side is mirrored. Eq. \ref{eq:map} is a functional form that ensures the derivative of $a$ with respect to $i$ to be zero on the sides and in the center. In Fig. \ref{fig:cavity-design} we show the bandstructure of the unit cell and classify the modes according to the symmetry of the tangential electric field with respect to two planes: $XY$ and $ZX$, with $x$ parallel to the beam.
\begin{figure*}[ht]
    \centering
    \includegraphics[width=0.80\linewidth]{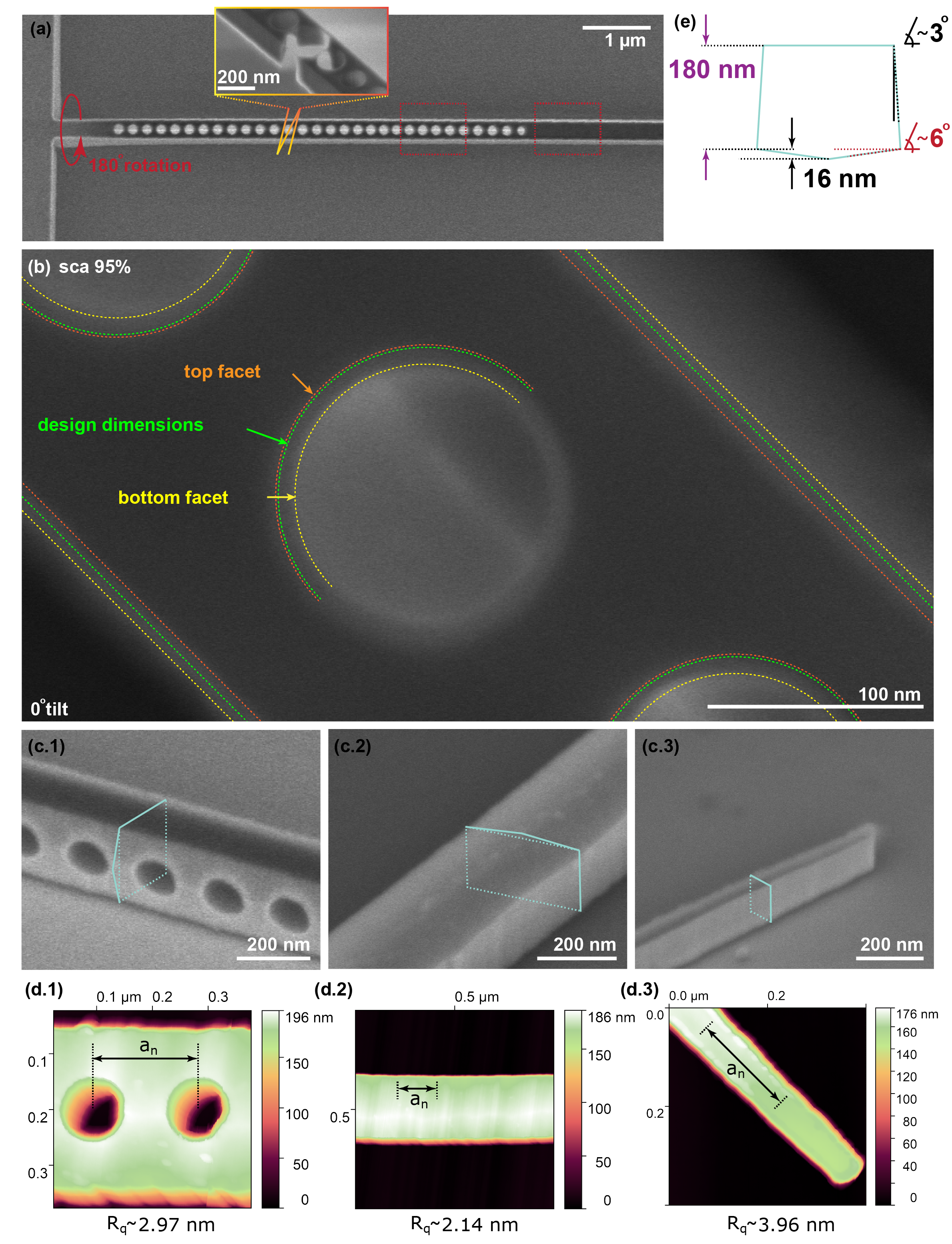}
    \caption{Scanning Electron Microscopy (SEM) and Atomic Force Microscopy (AFM) device geometry characterization on twin diamond samples (Sample 2 and Sample 3), representative examples of QIE-based fabrication process performance. \textbf{(a)} Typical diamond photonic crystal cavity; \textbf{(b)} High magnification SEM of a PCC transfer pattern on Sample 2; \textbf{(c)} SEM of device backside of (c.1) cavity region, (c.2) waveguide region, (c.3) taper-end region; \textbf{(d)} AFM topography analysis (Sample 3) of device bottom facet: cross-section and surface roughness \ce{R_q} (evaluated over \ce{a_n} = \SI{196}{\nano\meter}) of \textbf{(d.1)} cavity region, V-shaped triangular cross-section and \ce{R_q} = \SI{2.97}{\nano\meter}, \textbf{(d.2)} waveguide region, V-shaped triangular cross-section and \ce{R_q} = \SI{2.14}{\nano\meter}, \textbf{(d.3)} taper-end region, rectangular cross-section and \ce{R_q} = \SI{3.96}{\nano\meter}; \textbf{(e)} Estimated full cross-section of the device in cavity region from a combination of AFM topography and top-down high magnification SEM.}
    \label{PCC_fig2}
\end{figure*}

The devices are single-sided photonic crystal cavities (schematically illustrated in Fig. \ref{PCC_fig1}), consisting of a strong mirror (over \ce{M_s} mirror unit cells, blue-shaded), central cavity region (over 2N unit cells, red-shaded) and a weak mirror (over \ce{M_w} variable number of mirror holes, blue shaded). The design of the suspended devices is inspired from Ref. \cite{nguyen2019integrated}, where we maintain a constant value of the lattice constant \ce{a_n}, equal for both the strong and weak mirrors. Effectively, in the presented design the two mirrors have the same stop-band, with the weak mirror characterized by a reflectance dependent on the number of mirror unit cells $M_w$. The directionality in the out-coupling from the cavity to the waveguide is realized by reducing the number of mirror unit cells. Next to the the weak in-coupling mirror, the latter is adiabatically tapered to the waveguide region (green shaded, width $w_{wg}$ same as the cavity width) by appending additional L mirror unit cells (yellow shaded), linearly decreasing circular hole radius from optimal r to $r_{min, wg}$ ($\sim$\SI{25}{\nano\meter} feasible with the QIE-based fabrication process parameters), whereas the lattice constant $a_n$ is kept constant. Next, the waveguide width is reduced from $w_{wg}$ to end-taper width $w_{t}$, over $L_{t}$ taper length (cyan).

In order to guarantee structural stability, a careful trade-off between the overall length of free-hanging structure is considered, sweeping the coupling taper length. To allow for efficient both in-contact and lensed-fiber configuration coupling to the devices, the taper-end to diamond substrate distance $d_{ts}$ is fixed. On the overall diamond substrate, we sweep the cavity device design by varying the scaling factor on the lattice constant \ce{a_n}, waveguide width \ce{w_{wg}} and circular hole radius \textit{r} incrementally by $\Delta$=1\%, with respect to the nominal values (scaling factor 100\%) determined via finite element method optical simulations in COMSOL$\text{\textregistered}$. Thus we vary the scaling factor from 95\% to 110\% with respect to nominal 100\% device parameters. This strategy is adopted in order to preventively compensate cavity resonance shifts from introduced fabrication process variability and yield cavity devices with resonances $\lesssim$\SI{619}{\nano\meter} (SnV center emission), thus allowing for dynamic in-situ cavity resonance tuning to the embedded SnV centers emission lines. 

The chiplet device consists of an array of single-sided cavity devices with variying $M_w$=1 to 6, inter-device gap distance $g_n$ (center-to-center: \SI{3}{\micro\meter}) is decided as a trade-off between substrate space optimization as well as compatibility with the QIE undercut and upward etch, guaranteeing uniform device thickness across the chiplet. The devices are fixed to the diamond substrate via the end waveguide region distanced from the strong mirror $\sim$\SI{850}{\nano\meter} and are fully free-hanging.

\section{Fabrication Process}
\label{sec:fabrication-process}

\begin{table*}[ht]
\centering
\scriptsize
\begin{tabular}{>{\centering\arraybackslash}m{1.5cm}||>{\centering\arraybackslash}m{2cm}|>{\centering\arraybackslash}m{2cm}|>{\centering\arraybackslash}m{2cm}||>{\centering\arraybackslash}m{2cm}|>{\centering\arraybackslash}m{1cm}||>{\centering\arraybackslash}m{1.2cm}|>{\centering\arraybackslash}m{1.2cm}|>{\centering\arraybackslash}m{1.5cm}}

  & \multicolumn{3}{c||}{\textbf{Pre-anneal substrate preparation}} & \multicolumn{2}{c||}{\textbf{Sn implantation}} & \multicolumn{3}{c}{\textbf{SnV activation}} \\ \hline \hline
\textbf{Step} & \textbf{Clean} &\textbf{Strain relief}  & \textbf{Clean} &  &  &\textbf{Clean} & \textbf{Anneal} & \textbf{Clean} \\ \hline \hline
\textbf{Parameter}   & HF 40\% + 2x Piranha    & \ce{Ar/Cl_2}, \ce{O_2}  & Piranha  &Dose
$[\si{\text{ions}\per\centi\meter\squared}]$
 & Implant angle &Tri-acid & Vacuum anneal &Piranha + di-acid\\ \hline

\textbf{Value}      &20 min + 2x 20 min   &\SI{5}{\micro\meter} + \SI{6}{\micro\meter}   & 30 min  &\num{3e10}, \num{1e11} &\ang{7} &4 hours @ \SI{120}{\degreeCelsius} &4 hours @ \SI{1100}{\degreeCelsius}  &20 min @ \SI{80}{\degreeCelsius} + 4 hours @ \SI{190}{\degreeCelsius} \\ \hline

\end{tabular}
\caption{\textbf{SnV center creation main steps and parameters} | Left: Pre-anneal substrate preparation; | Center: Sn ion implantation parameters; Right: SnV center activation.}
\label{Ch_PCC_table_PCC_prefabSnVimplant}
\end{table*}

The devices presented in this work are fabricated in two main phases: SnV creation and nanophotonic device fabrication, respectively.
SnV centers are generated by ion implantation and high temperature annealing, for which the main parameters are summarized in Table \ref{Ch_PCC_table_PCC_prefabSnVimplant}. The sample fabrication process starts with pre-implantation surface treatment of a \{001\} surface-oriented electronic-grade diamond substrate (ElementSix). The sample substrate is first cleaned in a wet HF 40\% inorganic solution at room temperature for 20 minutes, followed by Piranha (ratio 3:1 of \ce{H2SO4 (95\%)}~:~\ce{H2O2 (31\%)}) inorganic solution for 20 min at \SI{80}{\celsius}. Next, the superficial $\sim \SI{5}{\um}$ etching via inductively-coupled-plasma reactive-ion-etching (ICP-RIE) \ce{Ar/Cl2} plasma chemistry recipe follows in order to remove the residual polishing-induced strain from the surface of the substrate. An additional $\sim$ \SI{6}{\um} ICP-RIE \ce{O2} chemistry plasma etch is performed in order to remove residual chlorine contamination from the previous etching step \cite{ruf_cavityenhanced_2021}. The sample is then inorganically cleaned in a Piranha solution (20 min at \SI{80}{\celsius}) and implanted with \ce{^{120}}Sn ions (implantation dose \num{3e10} and \num{1e11} ions/\ce{cm^{2}} with an implantation energy of \SI{350}{keV}, \ang{7} implantation angle). Prior to the activation of the SnV centers by high temperature low pressure annealing (\SI{1100}{\degreeCelsius} with surface-protected annealing method \cite{codreanu_diamond_2025}, a triacid cleaning (ratio 1:1:1 of \ce{HClO4}$(70 \%)$~:~\ce{HNO3}$(70\%)$~:~\ce{H2SO4}$(>99\%)$, at \SI{120}{\degreeCelsius}) is performed for 4 hours in order to remove any residual organic contamination, whereas post-annealing the sample undergoes a 30 minutes Piranha cleaning at \SI{80}{\degreeCelsius} and 4 hour di-acid cleaning (ratio 1:1 of \ce{HClO4}$(70 \%)$~:~\ce{H2SO4}$(>99\%)$, at \SI{190}{\degreeCelsius}) in order to remove any superficial graphite thin film layer (on non-covered edges of the diamond substrate), as well as to oxygen terminate the graphite-free overall substrate surface. To assess the successful activation of the SnV centers, the sample is characterized (prior to the nanofabrication of the suspended structures) at \qty{5}{\K} in a closed-cycle cryostat to identify the activated SnV centers (employing PL and PLE measurement sequences).

The nanofabrication of photonic crystal cavity (PCC) structures (developed in \cite{codreanu_diamond_2025}) follows the crystal-dependent quasi-isotropic-etch (QIE) undercut-based fabrication process developed in references \cite{khanaliloo_high-qv_2015, mitchell2019realizing, mouradian_rectangular_2017, wan_large-scale_2020, rugar_quantum_2021} and \cite{ruf_cavityenhanced_2021}. A schematic of this process is shown in \cite{codreanu_diamond_2025, pasini_nonlinear_2024}.

Specifically, we start with an extensive inorganic clean for 20 min in HF (40\%) at room temperature, followed by double Piranha (ratio 3:1 of \ce{H2SO4 (95\%)}~:~\ce{H2O2 (31\%)}) inorganic solution for 20 min at \SI{80}{\celsius} each.

Next, deposition and patterning of a hard mask material thin film layer of $\sim$ \SI{81}{nm} \ce{Si_xN_y} via inductively-coupled-plasma-chemical-vapour-deposition (ICPCVD). The PCC design is longitudinally aligned with the  $\langle 110\rangle $ diamond crystallographic orientation and exposed via e-beam lithography of $\sim$ 220 nm of AR-P-6200.09 positive tone resist. To avoid charging effects of the diamond substrate during the e-beam exposure, the surface of the e-beam resist is coated with $\sim40$ nm Electra 92 (AR-PC 5090) conductive polymer. Next, after e-beam exposure, this is removed by immersing the sample in 60 s gentle stirring in \ce{H2O} and \ce{N2} blow-dry (to dissolve Electra 92 and fully dry the residual \ce{H2O}). The resist is then developed in 60 s pentyl acetate, 5 s ortho-xylene, 60 s isopropyl alcohol (IPA). We then proceed with the transfer pattern into the \ce{Si_xN_y} hard mask material by means of ICP-RIE etch in a \ce{CHF3}/\ce{O2} based plasma chemistry.
Next, the complete removal of residual e-beam resist in a two-fold wet step follows: we first proceed with a coarse resist removal in a PRS 3000 positive resist stripper solution, followed by fine removal of resist residues in an extensive double Piranha inorganic clean. Such considerable inorganic cleaning is employed to prevent micro-masking within the next dry etch steps that can potentially be caused by organic residues on the sample. 

The pattern from the \ce{Si_xN_y} hard mask material is transferred in the diamond substrate by etching top-down in a dry ICP-RIE \ce{O2} anisotropic etch for an extent of $\sim$2.5x the designed thickness of the devices. It is crucial to note that the \ce{O2} dry etch heavily affects the aspect ratio of the patterned \ce{Si_xN_y} hard mask: we observe the etch rate of \ce{Si_xN_y} to be higher at the edges of the patterned structures compared to the determined etch rate of \ce{Si_xN_y} on flat area test samples, leading to enhanced erosion of the \ce{Si_xN_y} at the edges of the nanostructures. This yields rounded vertical sidewalls with few nanometers of hard mask material at the diamond to \ce{Si_xN_y} interface \cite{ruf_cavityenhanced_2021}. This leads to weak points across the top surface of the hard mask that can compromise the integrity of the hard mask in the downstream QIE-based process fabrication steps. We circumvent this challenge by carefully tuning the trade-off between sufficient anisotropic diamond etch and \ce{Si_xN_y} mask integrity, such that the latter withstands the following ICP-RIE dry etch steps foreseen by the overall QIE-based fabrication process. 

Next, conformal atomic layer deposition (ALD) of $\sim$ 25 nm of \ce{AlO_x} for hard mask coverage of devices vertical sidewalls and the horizontal coverage of \ce{AlO_x} is fully removed by ICP-RIE etch in a \ce{BCl3}/\ce{Cl2} plasma chemistry etch. This opens access to the diamond substrate: we directly follow with sequential QIE undercut and upward etch steps, interleaved by SEM monitoring. The full release and upward quasi-isotropic total etch time is considerably shorter (85 minutes) when compared to the \SI{65}{\degreeCelsius} QIE undercut recipe, demonstrated suitable for complete undercut of nanophotonic waveguides \cite{pasini_nonlinear_2024}. Finally, the hard mask materials are removed in extensive inorganic treatment for 20 min in HF (40\%) at room temperature.

The overall qualitative characterization of the fabrication steps presented in this work has been executed on a scanning electron microscope Hitachi SEM Regulus system. The ICPCVD deposition rates and ICP-RIE etch rates have been pre-characterized on additional test samples, in parallel to the fabrication of the diamond sample employed in this work following the same fabrication methods and parameters. Such etch tests have been conducted employing silicon substrate samples, with the thin films of interest deposited in parallel to the diamond substrate. Optical parameters and thickness of the employed materials have been determined via spectroscopic ellipsometry method on a Woollam M-2000 (XI-210) tool. The QIE rate has been characterized employing supplementary diamond test samples (fabrication parallel to the sample in this work) and analyzed via SEM inspection. Cross-section of obtained diamond devices has been determined from a combination between high magnification SEM (Hitachi SEM Regulus) inspection and atomic force microscopy (Bruker FastScan AFM) topographic analysis.

\section{Device Geometry Characterization}
\label{sec:geometry-characterization}
\begin{figure*}[ht]
    \centering
    \includegraphics[width=0.8\linewidth]{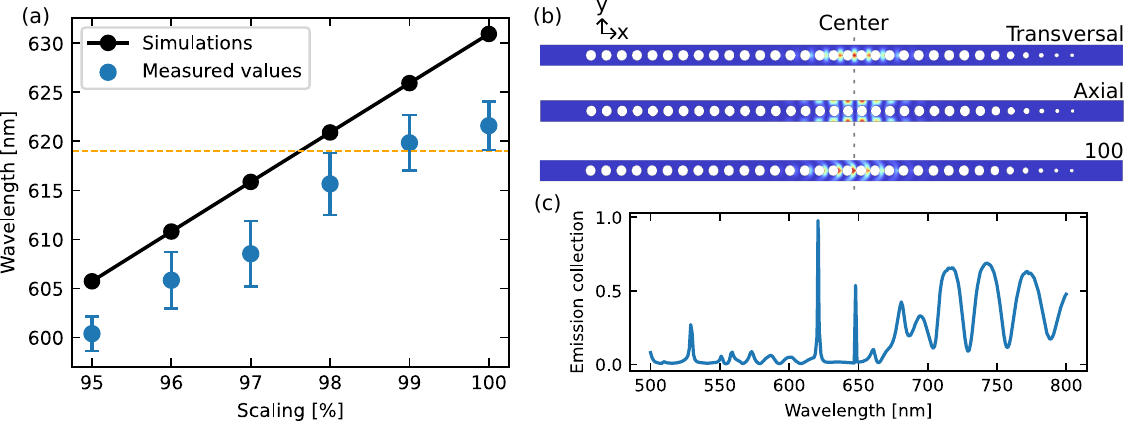}
    \caption{
    Cavity properties.
    \textbf{(a)}
    Resonance wavelengths as a function of the scaling parameter extracted from simulations (black) and measured statistical averages (blue).
    The orange dashed line lies at $619$~nm, the ZPL of the SnV center. \textbf{(b)}
    Cooperativity color map as a function of the emitter's position and the three dipole orientations.
    \textbf{(c)}
    Fraction of the emission collected in the waveguide from a y-polarized dipole centered in the cavity as a function of the emission wavelength.
    The two central peaks correspond to the fundamental and first-order cavity modes.
    }
    \label{fig:cavity-sim}
\end{figure*}

QIE-based fabrication process unavoidably introduces inaccuracy in the fabricated devices geometry, deviating from the simulated rectangular unit cell. This prevalently affects the hole radius $r$, device width $w_{wg}$ and device cross-section (as shown in Figure \ref{PCC_fig2}), inherently reducing quality factor and shifting the cavity resonance with respect to the design one. Here we present the quantitative analysis on typical consistent geometry bias and transversal cross-section, by combining Scanning Electron Microscopy (SEM) inspection and atomic Force Microscopy (AFM) topography analysis. The device geometry analysis is performed on two twin diamond samples (SEM geometry analysis on Sample 2, AFM topography analysis on Sample 3), fabricated within the same fabrication run and same fabrication process parameters as the sample used in this work (Sample 1).

Transfer pattern accuracy ($a_n$, $r$ and $w_{wg}$) yields differences on the order of few nanometers (not quantifiable via SEM analysis), whereas device thickness variability is $\sim$\SI{10}{\nano\meter} between the analyzed samples (Sample 2 and Sample 3 have a device thickness $\sim$\SI{10}{\nano\meter} lower than Sample 1).
The transversal device cross-section estimated in the following section (namely vertical sidewall angle deviation and bottom facet topography) is representative between the three fabricated diamond substrates.

High magnification SEM analysis of Sample 2 (Fig. \ref{PCC_fig2}(b), \ang{0} tilt angle view, pixel size $\sim$\SI{0.52}{\nano\meter}) enables inspection of fabricated devices top facet, allowing for accurate estimation of lattice constant $a_n$, hole radius $r$ and device width $w_{wg}$. In this example, the analysis is carried out on a PCC device with scaling 95 \% ($a_n$=0.95$\cdot$\SI{206}{\nano\meter}=\SI{195.7}{\nano\meter}, $r$=0.95$\cdot$\SI{71}{\nano\meter}=\SI{67.45}{\nano\meter}, $w_{wg}$=0.95$\cdot$\SI{275}{\nano\meter}=\SI{261.25}{\nano\meter}), that is the smallest device dimensions we pattern in our typical parameters scaling factor sweep.

Overlaying the CAD design (Fig.\ref{PCC_fig2} (b), dashed green) to the underlying corresponding scaling factor 95\% fabricated device SEM pictogram (pixel size \SI{0.53}{\nano\meter}), no significant deviation from the optimal $a_n$ can be observed, confirming a high accuracy of $a_n$ in transfer pattern. When comparing the circular hole diameter and device width on the top facet (dashed orange) with the design parameters (r and $w_{wg}$), an r systematic bias of $\sim$\SI{4}{\nano\meter} (radius larger than design) and a $w_{wg}$ systematic bias $\sim$\SI{20}{\nano\meter} (width smaller than design) can be identified.

Next, high magnification top-down SEM inspection allows for an estimation of the vertical sidewall angle. That is, in the case of design sidewall angle of \ang{90}, the top facet contour (Fig. \ref{PCC_fig2} (b), dashed orange) of etched devices should perfectly align with the bottom facet contour (dashed yellow). By measuring the horizontal extent between the top facet contours and bottom fact contour, we identify on average a distance of $\sim$\SI{10}{\nano\meter}, indicating that the sidewall angle is $\lesssim$\ang{90}. When measuring the same quantity on a scaling factor 100\% ($a_n$=\SI{206}{\nano\meter}, $r$=\SI{71}{\nano\meter}, $w_{wg}$=\SI{275}{\nano\meter}) PCC device, we find similar values, within the SEM pixel size resolution error, thus independent on the absolute parameter values, given a certain device scaling factor. A measurement of (same scaling factor 95\%) device thickness from SEM inspection (\ang{45} tilt angle view, pixel size \SI{1.98}{\nano\meter}, Figure \ref{PCC_fig2} (d)), yields a vertical extent from top facet to bottom facet of $\sim$\SI{188}{\nano\meter}. Assuming the bottom facet of the device characterized by a flat planar surface, the estimated sidewall angle deviation from the designed $\sim$\ang{90} is of $\sim$\ang{3}.

In the typical chiplet layout (Fig. \ref{PCC_fig2} (a)), free-hanging devices are attached to the bulk diamond substrate from the end-side of the strong mirror region of the device. On the diamond Sample 3, a device breaking mechanism takes place (shown in inset). By SEM inspecting the broken PCC structures, we are able to identify the parent chiplet. For devices repositioned to expose bottom facet, qualitative SEM inspection of the cavity region and waveguide region (Figure \ref{PCC_fig2} (c.1) and (c.2) respectively, on three separate broken devices from different parent chiplets across the sample) allows to SEM resolve a triangular cross-section and rectangular cross-section in the taper-end region (Figure \ref{PCC_fig2} (c.3)). At the same time, SEM resolved surface roughness can be identified, thus taking into account the pixel size ((c.1) \SI{3.17}{\nano\meter}, (c.2) \SI{1.98}{\nano\meter} and (c.3) \SI{1.44}{\nano\meter} pixel size respectively) in the shown pictograms, we suspect the surface roughness of the bottom facet to be of similar order of magnitude.

Next, AFM topography scan on Sample 3 (representative of typical fabrication imperfections) of a PCC device (scaling factor 98\%) exposing the topside facet enables a quantitative evaluation of the circular hole diameter and waveguide width. This allows to quantify the discrepancy between top-facet and design parameter values from topographic measurements. A consecutive AFM topography scan of a PCC device (scaling 97\%), exposing the bottom facet enables both a quantitative evaluation of bottom topography cross-section and surface roughness evaluation, as shown in Figure \ref{PCC_fig2} (d.1), (d.2) and (d.3), respectively. 

By comparing the top-side facet and bottom-side facet evaluated values of hole radius $r$ and waveguide width $w_{wg}$, the AFM analysis confirms the accuracy of the SEM analysis presented above, here measuring a bottom device facet $w_{wg}$ extent $\sim$\SI{20}{\nano\meter} higher and a $r$ extent $\sim$\SI{10}{\nano\meter} lower when comparing to the same parameters measured on the device top facet. Next, the topography maps of the cavity region and waveguide region show a triangular V-shape bottom facet, with a height of $\sim$\SI{16}{\nano\meter}, whereas the topography map of the taper-end displays a rectangular cross-section. 

Combining the evaluated experimentally measured device parameters allows us to infer the PCC device cavity and waveguide regions cross-section, where we estimate the vertical sidewall angle $\sim$\ang{3} and the bottom facet V-shaped with an angle $\sim$\ang{6}, schematically illustrated in Figure \ref{PCC_fig2} (e). The total thickness difference between the cavity and waveguide regions is $\sim$\SI{10}{\nano\meter} in the specific evaluated 97\% device scaling factor here presented. The observed device geometry reflects the nature of aspect-ratio QIE undercut dependency, for which the complete upward etch flattening of narrow devices (i.e. design taper-end width of $w_{t}$=\SI{50}{\nano\meter}) takes place earlier in the etch process, when compared to wider devices (i.e. design cavity and waveguide region width of $w_{t}$=\SI{275}{\nano\meter}).

\section{Device Statistics}
\label{sec:device-stats}
\begin{table}[h!]
    \caption{Cavity design parameters}
    \label{tab:cav_geom_params}
    \centering
    \begin{tabular}{|c|c|c|}
        \hline
        Parameter & Sample 1 & Sample 2\\
        \hline
        \hline
        Investigated & 232  & 254 \\
        \hline
        With Resonance & 200 & 216 \\
        \hline
        Visually Broken & 22 & 25 \\
        \hline
        No Resonance & 10 & 13 \\
        \hline
        \hline
        Q Measured  & 163 & 164 \\
        \hline
        Not in Reflection & 27 & 7 \\
        \hline
        Spectrometer Range & 7 & 0 \\
        \hline
        Spectrometer Resolution & 0 & 39 \\
        \hline
        Fit Error $> 20\%$ & 3 & 8 \\
        \hline
    \end{tabular}
\end{table}

The reported quality factor statistics exclude several devices from the investigated subset on each fabricated diamond sample. 
For Sample 1, we report a total of 163 quality factors, excluding devices where the fundamental cavity mode falls outside the range of our spectrometer (7) or where we do observe a cavity resonance but the measurement contrast in reflection is too low to attribute a quality factor (27).
For these cavities one can measure a resonance when the fiber is placed on top of the cavity.
However, this influnces the quality factor, so we do not attribute a value for $Q$.

For Sample 2, there are (39) cavities where the first higher order mode can be observed and the fundamental mode should fall within the spectrometer range
(the distance between fundamental mode and first higher order mode is $\sim \SI{30}{\nano\meter}$).
However, no fundamental mode is observed.
We attribute this to the limited spectometer resolution.

\section{Cavity Simulations}
\label{sec:cavity-simulations}
\begin{figure*}[ht]
    \centering
    \includegraphics[width=\linewidth]{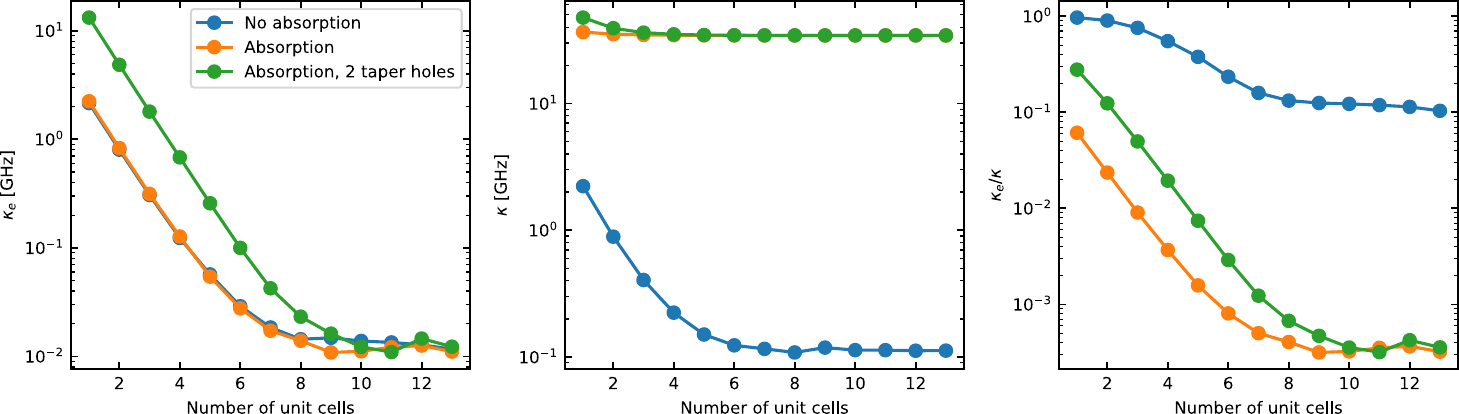}
    \caption{
    Directional decay rates and waveguide coupling efficiencies. 
    \textbf{(a)} Waveguide decay rates $\kappa_e$, \textbf{(b)} total decay rates $\kappa$ and \textbf{(c)} waveguide coupling efficiencies $\kappa_e/\kappa$ as a function of the number of unit cells in the weak mirror region. 
    The three curves correspond to a cavity without absorption and all the unit cells in the taper region (blue), a cavity with absorption and all the unit cells in the taper region (orange), and a cavity with absorption and only $2$ unit cells in the taper region (green).
    We used an imaginary component for the refractive index of $10^{-4}$.
    }
    \label{fig:kappa-sim}
\end{figure*}
The fabrication process introduces systematic imperfections in the geometry, namely, deviations in the width of the nanobeam, sidewall angles, and non-flat bottom surfaces.
Therefore, the cavity design with the nominal parameters presented in the previous section is not expected to fully reproduce the measured characteristics of the cavities.
These imperfections are compensated by fabricating devices with a smaller thickness (from $230$~nm to $188$~nm) and introducing a scaling parameter.
The latter is a scaling factor that is applied to all geometrical parameters, except for thickness, and is swept on the fabricated sample.

After measuring the device cross-section via SEM and AFM, we incorporate it into our Finite Element (FEM) simulations with COMSOL Multiphysics \cite{comsol}. 
We find a disagreement of $\sim6$~nm between our simulations and the average cavity resonances measured at room temperature, see Fig. \refsub{fig:cavity-sim}{a}.

Based on the simulated electric fields, we plot the cooperativity as a function of the emitter position and dipole orientation in Fig \refsub{fig:cavity-sim}{b}. These figures illustrate the importance of deterministic positioning of the color centers to achieve the maximum possible cooperativities. For the latter, we used the following definition \cite{reiserer_cavitybased_2015, herrmann_coherent_2024}.
\begin{equation}
    C_\text{bulk} \equiv \frac{\gamma_\s{cavity}}{\gamma_\text{bulk}}= \frac{3}{4\pi ^2}\frac{Q}{\frac{V}{(\lambda/n)^3}}|\hat{d}_i\cdot \hat{E}(\vec{x})|^2\cdot u^2(\vec{x}) \cdot \beta_0,
    \label{eq:coop}
\end{equation}
where $Q$ is the quality factor of the cavity mode, $V\equiv \frac{\int{\epsilon (\vec{x}) |E|^2dV}}{\max\{\epsilon (\vec{x})|\vec{E}(\vec{x})|^2\}}$ is the mode volume, $\lambda$ is the vacuum wavelength of the emitter and cavity at resonance, $n$ is the refractive index of diamond, $\hat{E}(\vec{x})$ is the unit vector for the electric field at the emitter's location $\vec{x}$, $u^2(\vec{x})\equiv \frac{\epsilon (\vec{x})|\vec{E}(\vec{x})|^2}{\max\{\epsilon (\vec{x})|\vec{E}(\vec{x})|^2\}}$, and $\beta_0$ is the fraction of decay from the excited state into the line resonant with the cavity. 

The orientation of the electric dipole of the emitter is along the $\langle 111\rangle $ crystallographic axis of diamond and is denoted by $\hat{d}_i$.
We consider three dipole orientations $i\in \{\text{Transversal}, \text{Axial}, \langle 100\rangle\}$, where
\begin{align}
    \hat{d}_\text{transversal} &= \frac{1}{\sqrt{3}}\left(0, \sqrt{2}, 1\right),\\
    \hat{d}_\text{axial} &= \frac{1}{\sqrt{3}}\left(\sqrt{2}, 0, 1\right),\\
    \hat{d}_{\langle 100\rangle} &= \frac{1}{\sqrt{3}}\left(1, 1, 1\right),
\end{align}
which are defined with respect to the coordinate axes in Fig \refsub{fig:cavity-sim}{b}.
The devices in the main text are fabricated along the $\langle 110\rangle $ crystallographic axis of diamond, so only the transversal and axial orientations are possible. The remaining dipole orientation considered would correspond to a device fabricated along the $\langle 100\rangle$ crystallographic axis.

\begin{figure*}[ht]
    \centering
    \includegraphics[width=\linewidth]{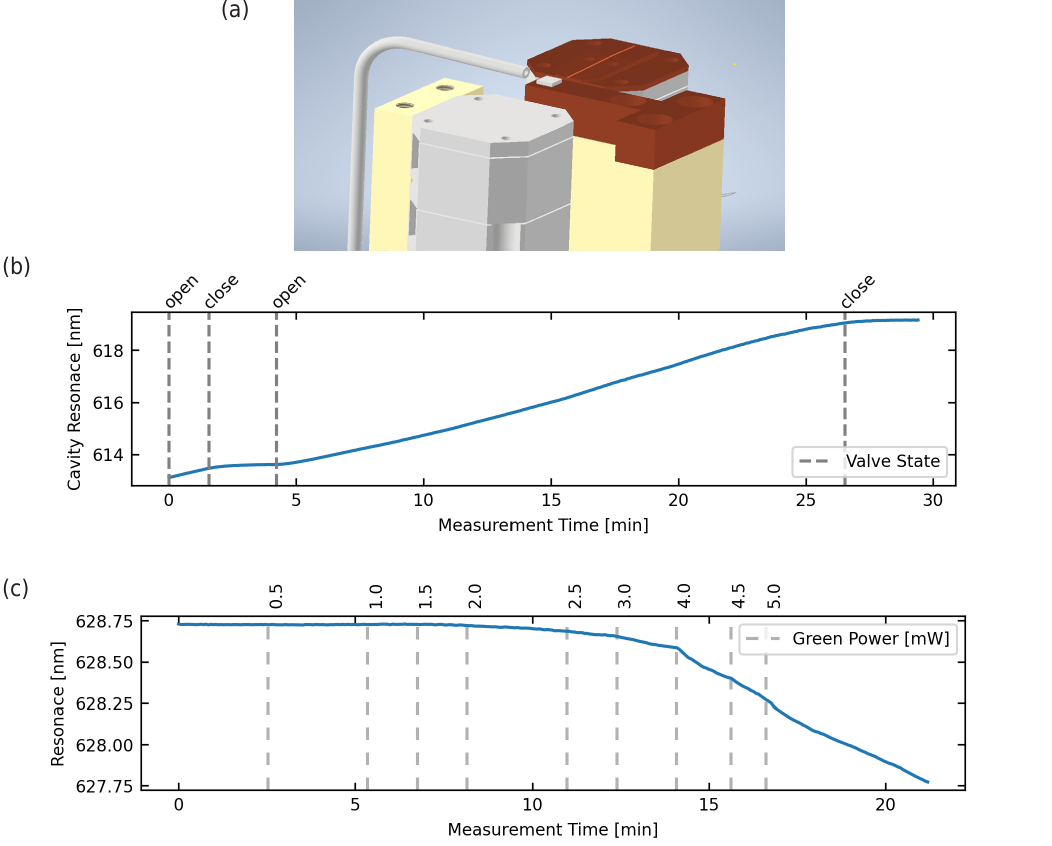}
    \caption{
    Gas tuning.
    \textbf{(a)}
    Render of the inside of the cryostat.
    The diamond sample is glued on a copper cold finger with silver paste.
    A $\s N_2$ gas line is pointed at the sample.
    The fiber is stuck to a machined copper holder with tape and can be moved with a piezo positioner.
    The second set of positioners is not used in this work.
    \textbf{(b)}
    The PCC is tuned forward by opening the leak valve 
    (flow rate $\sim \SI{1e-9}{\milli\bar\liter\per\second}$).
    \textbf{(c)}
    The PCC is tuned backward by shining a green laser onto the PCC.
    The speed of tuning can be adjusted by the laser power.
    }
    \label{fig:gas-tuning}
\end{figure*}

Note that this definition of cooperativity compares the rate of emission into the cavity at resonance with the spontaneous decay rate of the emitter in bulk diamond.
However, in our case, the emitter is in a nanophotonic structure.
A more accurate definition of cooperativity that is found experimentally would be $C\equiv \frac{\gamma^\prime (\Delta = 0)-\gamma^\prime (\Delta \to \infty)}{\gamma^\prime (\Delta \to \infty)}$, which simply compares the decay rates of the emitter between the off and on resonance cases. 
The latter is the cooperativity reported in the main text.
The lifetime of an emitter in a nanophotonic structure is expected to be larger with the proximity of the emitter to a vacuum interface \cite{gorlitz_spectroscopic_2020}.
Therefore, the definition of Eq. \ref{eq:coop} is a lower bound for the attainable cooperativities.

Additionally, we simulate a y-polarized electric dipole in the center of the cavity and compute the fraction of power that exits through the coupling waveguide mode.
In Fig. \refsub{fig:cavity-sim}{c} we plot this transmission as a function of the dipole's frequency.
We observe a high transmission when the dipole is on resonance with the two cavity modes, but also at longer wavelengths.
Therefore, both the ZPL and part of the PSB of the emitters can be collected through the waveguide.

We now investigate the directional decay rates of the cavity.
The total cavity decay rate $\kappa = \omega/Q$ can be decomposed into two decay rates $\kappa=\kappa_e+\kappa_i$, where $\kappa_e$ is the decay rate into the waveguide and $\kappa_i$ is the intrinsic decay rate, which includes all other decay channels.
The simulations naturally provide the total decay rate $\kappa$ from the solution to an eigenvalue equation. 
We then compute the ratio between the power exiting through the waveguide and the power exiting through the entire domain.
From the equality $P_\text{waveguide}/P_\text{domain}=\kappa_e/\kappa$, we can extract $\kappa_e$.

Starting from a device with a simulated $Q\sim 4\cdot 10^6$, Fig. \refsub{fig:kappa-sim}{c} shows that the coupling factor into the waveguide $\kappa_e/\kappa$ goes to $\sim 1$ (blue curve) as the number of unit cells in the weak mirror region is lowered. 
However, the second device that we measured, which has $1$ unit cell in the weak mirror region, only achieves a coupling factor $\kappa_e / \kappa = 0.31$, see Sec. \ref{sec:reflection}.
The discrepancy between the simulated and measured value arises because the intrinsic decay rate $\kappa_i$ of the fabricated devices is much faster ($Q\sim 10^4$, see Table \ref{tab:cqed}) due to additional scattering losses, such as bottom surface roughness, whereas the decay rate into the waveguide $\kappa_e$ remains roughly equal, thereby lowering the coupling efficiency $\kappa_e/(\kappa_e+\kappa_i)$.

To demonstrate that $\kappa_e$ is roughly independent of these scattering losses, we artificially introduce absorption into the cavity by adding an imaginary part to the refractive index of the diamond that makes up the cavity.
The formula to extract $\kappa_e$ becomes $P_\text{waveguide}/(P_\text{domain}+P_\text{absorption})=\kappa_e/\kappa$.
In Fig. \refsub{fig:kappa-sim}{a} we see that $\kappa_e$ remains the same (orange curve), whereas $\kappa$ increases as expected, see \refsub{fig:kappa-sim}{b}.
In this case, we obtain a coupling factor  $\kappa_e / \kappa\sim 0.06$ for a weak mirror of 1 unit cell, which underestimates the measured coupling factor.

The number of unit cells in the taper region is $7$ by design.
However, most devices on the sample only have the largest $2$, the rest did not open.
We remove these from the simulation, and find that the decay rate into the waveguide increases (green curves), resulting in a coupling factor of $\kappa_e / \kappa\sim 0.28$ for a weak mirror of 1 unit cell.
This is probably because the taper region also acts like a mirror and reducing the number of unit cells further weakens the mirror.

The simulated device with absorption and $2$ unit cells in the taper region has a $Q\sim 10^4$ at $1$ unit cell in the weak mirror region, similar to device 2, see Table \ref{tab:cqed}. 
We further find a simulated $\kappa_e\sim 13$~ GHz that is in good order-of-magnitude agreement with the measured value $\kappa_e\sim 8$~GHz. 
Thus, in order to fabricate overcoupled devices, the designed $\kappa_e$ should be greater than the intrinsic decay rate $\kappa_i$ the fabricated devices are expected to have.

\section{Gas Tuning}
\label{sec:gas-tuning}

We deposit nitrogen gas onto the PCC to tune its resonance.
This technique, originally reported by Mosor et al. \cite{mosor_scanning_2005}, has been widely used 
\cite{strauf_frequency_2006,srinivasan_optical_2007,dalacu_scalable_2010,evans_photonmediated_2018,lettner_controlling_2024}.
The deposited nitrogen has a refractive index $> 1$ and modifies the evanescent field of the cavity.
Intuitively, the PCC with deposited gas has a larger lattice constant and is thus resonant for larger wavelengths.
The nitrogen (99.999\% purity) flow is controlled by a leak valve (Nenion, UHV-Leakvalve ND 3).
It is guided by a 1/16" inner diameter stainless steel tube assembly (Quantum Design, MI-4100-1184) directly onto the diamond sample (Fig. \refsub{fig:gas-tuning}{a}).
The flow rate is set to 
$\sim \SI{1e-9}{\milli\bar\liter\per\second}$.
This results in a tuning rate of about 
$\SI{0.2}{\nano\meter\per\minute}$ (Fig. \refsub{fig:gas-tuning}{b}).
After closing the valve, the resonance continues to tune for a small amount, presumably due to a trace amount of gaseous nitrogen that continues to deposit.
To tune the cavity backwards, we shine the green laser onto the defect region (Fig. \refsub{fig:gas-tuning}{c}).
This allows a precise tuning of the cavity resonance.

\section{C-QED Modelling}
\label{sec:cqed_modelling}
\begin{figure}[ht]
    \centering
    \includegraphics[width=\linewidth]{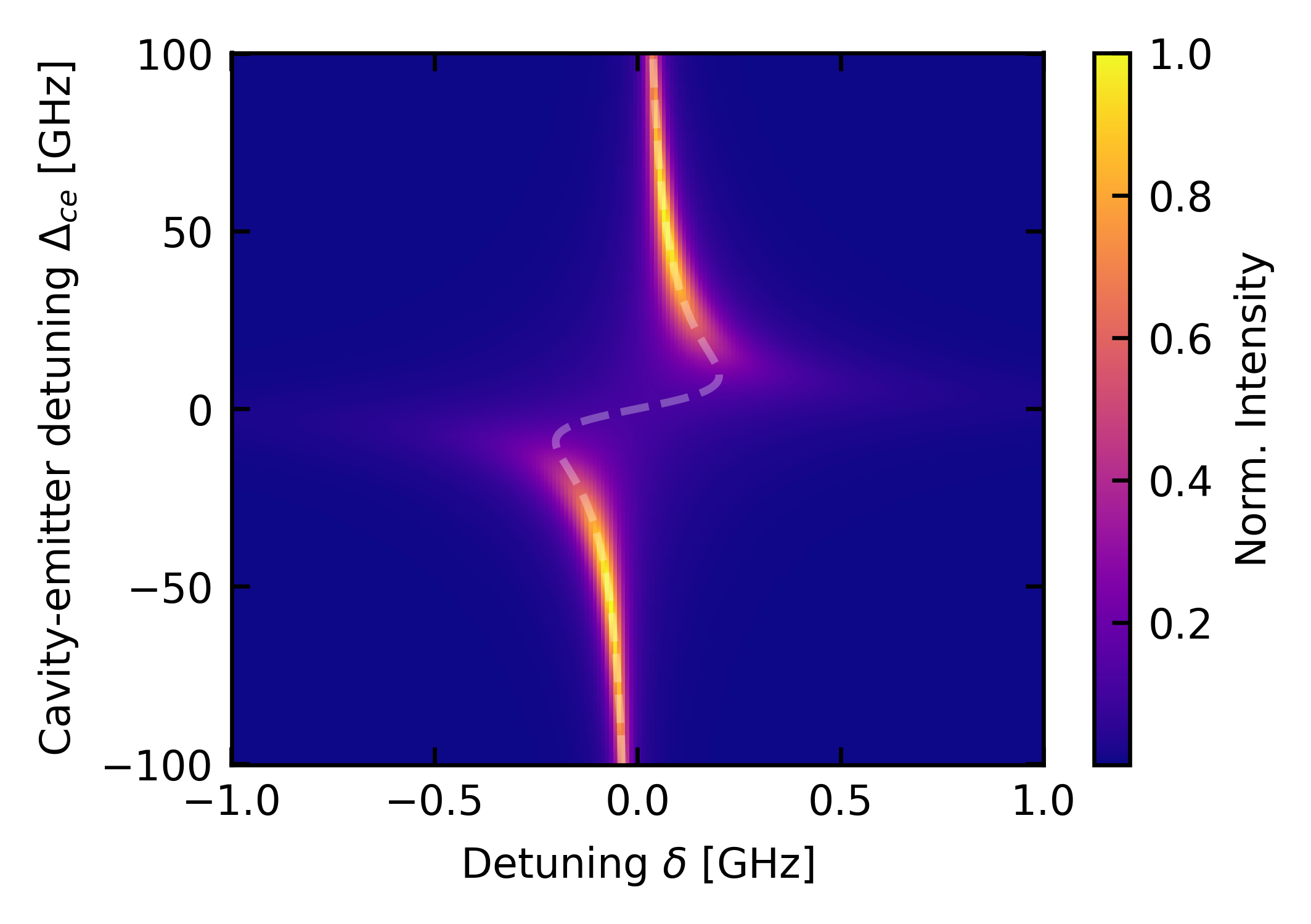}
    \caption{
    Normalized scattering intensity $|s|^2$ from Eq. \ref{eq:amplitude} as a function of laser-emitter detuning $\delta$ and cavity-emitter detuning $\Delta_\s{ce}$. The dashed line corresponds to an approximation of the frequency of brightest emission as given by Eq. \ref{eq:approx_deltamax}. For this plot we use $\{g,\kappa,\gamma_0\}/2\pi=\{1.94, 18.7, 0.02437\}$~GHz.
    }
    \label{fig:scattering}
\end{figure}

\begin{figure*}[ht]
    \centering
    \includegraphics[width=\linewidth]{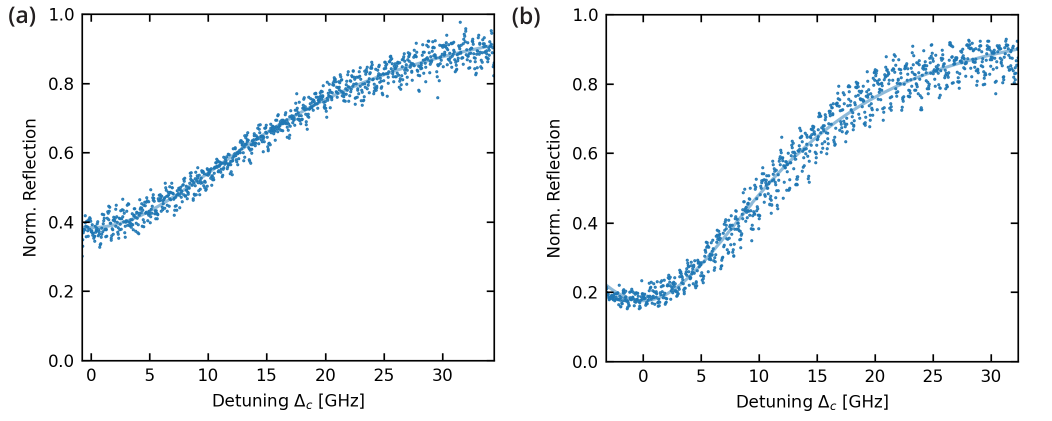}
    \caption{
    The cavity reflection is measured for device one \textbf{(a)} and device \textbf{(b)}.
    The data is fit with
    $R(\Delta_c) = 
    \left|
    1 - \frac{\kappa_e}{\kappa/2 + i\Delta_c}
    \right|^2$
    to obtain the outcoupling ratio $\kappa_e/\kappa = 0.21$ for Cavity 1 and $\kappa_e/\kappa = 0.31$ for Cavity 2.
    }
    \label{fig:reflection}
\end{figure*}

At our sample temperature and in the absence of a magnetic field, we can treat the SnV center as a two-level emitter coupled to a single-mode cavity. The system's dynamics is governed by the Jaynes-Cummings hamiltonian
% \begin{equation}
%     H_\text{JC}= \frac{\hbar}{2}\delta \sigma_z+\hbar \Delta a^\dagger a+\hbar g (a\sigma_\text{ge}^\dagger+a^\dagger\sigma_\text{ge})+i\hbar \epsilon(a^\dagger - a),
% \end{equation}
\begin{equation}
    H_\text{JC}= \frac{\hbar}{2}\delta \sigma_z+\hbar \Delta a^\dagger a+\hbar g (a\sigma_\text{ge}^\dagger+a^\dagger\sigma_\text{ge}),
\end{equation}
which is given in a frame rotating with the incident laser's frequency $\omega$. Here, $\sigma_\text{ge}=|g\rangle\langle e|$, $\sigma_z=|e\rangle\langle e|-|g\rangle\langle g|$, and $a$ is the cavity mode annihilation operator. The detunings are defined as $\delta = \omega_a-\omega$, $\Delta = \omega_c-\omega$, with $\omega_a$, $\omega_c$ the frequencies of the emitter and cavity, respectively. The cavity-emitter coupling is denoted by $g$.%, and $\epsilon $ is a parameter describing the cavity excitation strength.

% Both the cavity and the emitter couple to radiation modes which can be described by the collapse operators
Both the cavity and the emitter interact with bosonic baths. These bosonic baths describe the spontaneous emission of the emitter with a rate $\gamma_0$, and the different decay channels of the cavity $\kappa = \kappa_\text{in}+\kappa_l$, with $\kappa$ the total decay rate, $\kappa_\text{in}$ the rate into the port through which the cavity is excited and $\kappa_l$ the rate into the remaining loss channels.

% \begin{equation}
%     L_1=\sqrt{\gamma} \sigma_\text{ge},\qquad L_2=\sqrt{\kappa}a.
% \end{equation}
The quantum Langevin equations describe the evolution of system operators in the interaction frame and are given by \cite{gardiner_input_1985, tiecke_nanophotonic_2014}
\begin{subequations}
\begin{align}
    \dot{a}(t)=&
    -i\left[\Delta a(t)+g\sigma_{\text{ge}}(t)\right]-
    \frac{\kappa}{2}a(t) \nonumber\\
    &-\left[\sqrt{\kappa_{\text{in}}}a_{\text{in}}(t)+\sqrt{\kappa_l}\xi_l(t)\right], \label{eq:langevin_a}\\
    \dot{\sigma}_{\text{ge}}(t)=&
    -i\left[\delta\sigma_{\text{ge}}(t)-g\sigma_z(t)a(t)\right]-\frac{\gamma_0}{2}\sigma_{\text{ge}}(t) \nonumber\\
    &+\sqrt{\gamma_0}\sigma_z(t)\xi_\gamma(t),\label{eq:sigmage_langevin}\\
    \dot{\sigma}_z(t)=&
    -2ig\left[\sigma_{\text{ge}}^\dagger(t)a(t)
    -a^\dagger(t)\sigma_{\text{ge}}(t)\right]\nonumber\\
    &-2\gamma_0\sigma_{\text{ge}}^\dagger(t)\sigma_{\text{ge}} (t) \nonumber\\
    &-2\sqrt{\gamma_0}\left[\xi_\gamma^\dagger(t)\sigma_{\text{ge}}(t)+\sigma_{\text{ge}}^\dagger(t)\xi_\gamma(t)\right],\label{eq:sigmaz_langevin}\\
    a_{\text{out}}(t)&=a_{\text{in}}(t)+\sqrt{\kappa_\text{in}}a(t),\\
    b_\s{out}(t)&=\sqrt{\kappa_\s{out}}a(t),\\
    s_\s{out}(t)&=\sqrt{\gamma_0}\sigma_\s{ge}(t).
    \label{eq:input_output}
\end{align}
\end{subequations}
Here, $a_{\text{in}}(t)$ and $a_{\text{out}}(t)$ are the input-output field operators that represent the coupling between the cavity and the input channel (sometimes the reflection channel), $b_\s{out}$ is the output field operator at the transmission channel, $s_\s{out}$ is the output field operator representing spontaneous emission, and $\xi_\gamma(t)$ and $\xi_l(t)$ are input noise operators for the emitter and the cavity.

\begin{figure*}[ht]
    \centering
    \includegraphics[width=\linewidth]{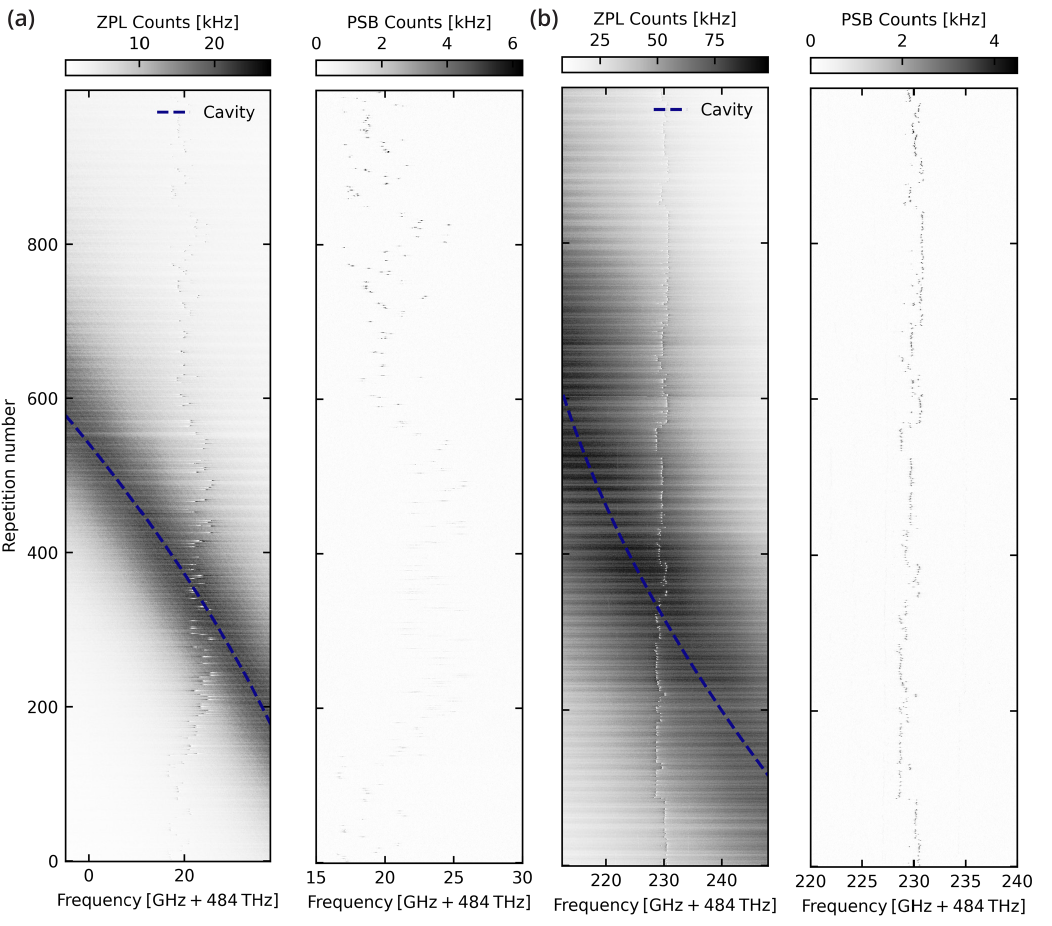}
    \caption{
    Complete set of transmission measurements for Cavity 1 \textbf{(a)} and Cavity 2 \textbf{(b)}.
    The cavity blue detunes slowly over time due to a small leak in the cryostat.
    }
    \label{fig:raw-data}
\end{figure*}

In the weak excitation regime (i.e., $\langle \sigma_z \rangle\simeq -1$ and $a\sigma_z=-a$) \cite{tiecke_nanophotonic_2014} and in the steady state ($\langle \dot{a}\rangle=\langle \dot{\sigma_\s{ge}}\rangle=0$) we find
\begin{subequations}
 \label{eq:steady_state}
 \begin{align}
    \langle a\rangle &= \frac{-\sqrt{\kappa_\s{in}}\langle a_\s{in}\rangle }{i\Delta+\frac{\kappa}{2}+\frac{g^2}{i\delta +\frac{\gamma_0}{2}}},
    \label{eq:steady_state0}\\
    \langle \sigma_\s{ge}\rangle &=\frac{-ig\langle a\rangle}{i\delta +\frac{\gamma_0}{2}}.
 \end{align}
\end{subequations}
In the absence of dephasing, it can be proven that the magnitudes of the reflection, transmission and scattering from the system can be simplified to \cite{julsgaard_reflectivity_2012}
\begin{subequations}
\begin{align}
    R &\equiv\frac{\langle a^\dagger_\text{out}a_\text{out}\rangle}{\langle a^\dagger_\text{in} a_\text{in}\rangle} = \left|\frac{\langle  a_\s{out}\rangle}{\langle  a_\s{in}\rangle}\right|^2,\\
    T &\equiv \frac{\langle b^\dagger_\text{out}b_\text{out}\rangle}{\langle a^\dagger_\text{in} a_\text{in}\rangle} = \left|\frac{\langle  b_\s{out}\rangle}{\langle  a_\s{in}\rangle}\right|^2,\\
    S &\equiv \frac{\langle s^\dagger_\text{out}s_\text{out}\rangle}{\langle a^\dagger_\text{in} a_\text{in}\rangle} = \left|\frac{\langle  s_\s{out}\rangle}{\langle  a_\s{in}\rangle}\right|^2.
\end{align}
\end{subequations}
From Eqs. \ref{eq:steady_state}, the reflection, transmission and scattering amplitudes take the form
\begin{subequations}
\begin{align}
    r&\equiv\frac{\langle  a_\s{out}\rangle}{\langle  a_\s{in}\rangle}=1-\frac{\kappa_\s{in}}{i\Delta+\frac{\kappa}{2}+\frac{g^2}{i\delta +\frac{\gamma_0}{2}}},\\
    t&\equiv\frac{\langle  b_\s{out}\rangle}{\langle  a_\s{in}\rangle}=-\frac{\sqrt{\kappa_\s{in}\kappa_\s{out}}}{i\Delta+\frac{\kappa}{2}+\frac{g^2}{i\delta +\frac{\gamma_0}{2}}},\\
    s&\equiv\frac{\langle  s_\s{out}\rangle}{\langle  a_\s{in}\rangle}=\frac{ig\sqrt{\kappa_\s{in}\gamma_0}}{(i\Delta+\frac{\kappa}{2})(i\delta +\frac{\gamma_0}{2})+g^2}.
    \label{eq:amplitude}
\end{align}
\end{subequations}
% From the input-output relations \cite{gardiner_input_1985, sipahigil_integrated_2016},
% the transmission through the cavity is given by
% \begin{equation}
%     T = \frac{\langle a^\dagger_\text{out}a_\text{out}\rangle}{\langle a^\dagger_\text{in} a_\text{in}\rangle} = \frac{\kappa_e \langle a^\dagger a\rangle}{\frac{|\epsilon|^2}{\kappa_\text{in}}}.
% \end{equation}
% Here, $\langle a^\dagger_\text{out}a_\text{out}\rangle$ and $\langle a^\dagger_\text{in}a_\text{in}\rangle $ denote the flux of photons out of and into the cavity, $\kappa_e$ is the coupling rate into the output waveguide mode and $\kappa_{in}$ is the coupling rate into the input free space mode.
The scattering intensity $S(\delta)$ presents a maximum approximately at
\begin{equation}
    \delta_\s{max} = \frac{g^2\Delta_\s{ce}}{\Delta_\s{ce}^2+\kappa^2/4}.
    \label{eq:approx_deltamax}
\end{equation}
The accuracy of the approximation decreases the higher the cooperativity. In Fig. \ref{fig:scattering} we plot the scattering intensity both as a function of $\delta$ and $\Delta_\s{ce}$. We see that the peak intensity decreases as the cavity goes into resonance with the emitter. We use this peak intensity envelope function as the fit in Fig. \refsub{fig:4}{d}. The variation in the frequency of the peak intensity cannot be observed in the raw data \refsub{fig:raw-data}{a} due to the inherent spectral diffusion and jumps of the emitter.

Without making assumptions about the emitter excitation regime, but assuming a large total cavity decay rate $\kappa$, it can be shown that \cite{carmichael, tiecke_nanophotonic_2014}
\begin{subequations}
\begin{align}
    \dot{\langle \sigma_\text{ge}\rangle}=
    &-\frac{\gamma_0}{2}\left(1+i\frac{2\delta}{\gamma_0}+\frac{C}{1+i\frac{2\Delta}{\kappa}}\right)\langle\sigma_\text{ge}\rangle\nonumber\\
    &-i\frac{\gamma_0}{2}\frac{Y}{1+i\frac{2\Delta}{\kappa}}\langle \sigma_\text{z}\rangle,\\
    \langle \dot{\sigma_z}\rangle =
    &-\gamma_0\left(1+\frac{C}{1+\left(\frac{2\Delta}{\kappa}\right)^2}\right)\left(1+\langle \sigma_z\rangle\right)\nonumber\\
    &+i\gamma_0\left(\frac{Y}{1+i\frac{2\Delta}{\kappa}}\langle\sigma^\dagger_\s{ge}\rangle-\frac{Y^*}{1-i\frac{2\Delta}{\kappa}}\langle\sigma_\s{ge}\rangle\right),
\end{align}
\end{subequations}
where $C=\frac{4g^2}{\kappa \gamma_0}$ and $Y=4g\sqrt{\kappa_\s{in}}\langle a_\s{in}\rangle/(\kappa \gamma_0)$. These are the optical Bloch equations of a two-level emitter but with different decay and excitation rates due to the interaction with the cavity. In particular, we see that when the emitter is in resonance with the laser (i.e., $\delta = 0$), the total decay rate $\gamma^\prime_0$ of the excited state population follows a Lorentzian with respect to the cavity-emitter detuning
\begin{equation}
    \gamma^\prime_0=\left(Cf(\Delta_\s{ce})+1\right)\gamma_0,
\end{equation}
where $f(\Delta_\s{ce}) = 1 / (1 + 4 \Delta_\s{ce} ^2/\kappa^2)$.
This is typically called Purcell enhancement.

\section{Reflection Measurements}
\label{sec:reflection}
\begin{table*}[ht]
    \caption{C-QED Parameters}
    \label{tab:cqed}
    \centering
    \begin{tabular}{|c|c|c|c|c|}
    \hline
    \multicolumn{5}{|c|}{\textbf{Cavity Parameters}} \\
    \hline
    Parameter & Symbol & Cavity 1 & Cavity 2 & Source \\
    \hline
    \hline
    Cavity Decay Rate & $\kappa$ &  $2\pi \times (19.0 \pm 0.3)$\,GHz & $2\pi \times (50.0 \pm 4.1)$\,GHz & Figs. \refsub{fig:4}{c} and \refsub{fig:deice-two}{a} \\
    \hline
    Cavity Decay Rate & $\kappa^\prime$ &  $2\pi \times (28.6 \pm 0.4)$\,GHz & $2\pi \times (75.0 \pm 0.4)$\,GHz & Figs. \refsub{fig:3}{c} and \refsub{fig:deice-two}{e} \\
    \hline
    External Decay Rate & $\kappa_e$ &  $2\pi \times (3.99 \pm 0.08)$\,GHz & $2\pi \times (16 \pm 1)$\,GHz & $\kappa_e / \kappa \times \kappa$ \\
    \hline
    External coupling & $\kappa_e / \kappa$ & $0.210 \pm 0.003$ & $0.310 \pm 0.002$ & Fig. \ref{fig:reflection} \\
    \hline
    Weak Mirror Unit Cells  & $M_w$ & $4$ & $1$ & Design \\
    \hline
    Quality Factor & $Q$ & $(25.40 \pm 0.37) \times 10^3$ & $(10.20 \pm 0.44) \times 10^3$ & $\omega / \kappa$ \\
    \hline
    Mode Volume & $V n^3 / \lambda^3$ & \multicolumn{2}{c|}{  0.45  } & FEM Simulations \\
    \hline
    Purcell Factor & $F_\s{P}$ & $(4.200 \pm 0.061)\times 10^3$ & $(1.60 \pm 0.13)\times 10^3$ & $\frac{3}{4\pi^2}\frac{Q}{Vn^3/\lambda^3}$ \\
    \hline
    \hline 
    \multicolumn{5}{|c|}{\textbf{Emitter Parameters}} \\
    \hline
    \hline
    Lifetime & $\tau_0$ & $(6.5 \pm 0.2)$\,ns & $(8.8 \pm 0.3)$\,ns & Figs. \refsub{fig:3}{c} and \refsub{fig:deice-two}{e} \\
    \hline
    Natural linewidth & $\gamma_0$ & $2\pi \times (24.5 \pm 0.8)$\,MHz & $2\pi \times(18.0 \pm 0.6)$\,MHz & Fig. \refsub{fig:3}{c} and \refsub{fig:deice-two}{e}\\
    \hline
    Linewidth & $\gamma$ & $2\pi \times (97.5 \pm 4.4)$\,MHz & $2\pi \times(52 \pm 5)$\,MHz & Figs. \refsub{fig:4}{d} and off-resonant linescans\\
    \hline
    Quantum Efficiency & $\eta_\s Q$ & \multicolumn{2}{c|}{  0.8  }  & \cite{iwasaki_tinvacancy_2017} \\
    \hline
    Debye-Waller Factor & $\eta_\s{DW}$ & \multicolumn{2}{c|}{  0.57  } & \cite{gorlitz_spectroscopic_2020} \\
    \hline
    Branching Ratio & $\eta_\s{BR}$ & \multicolumn{2}{c|}{  0.8  } & \cite{rugar_quantum_2021} \\
    \hline
    Beta Zero & $\beta_0$ & \multicolumn{2}{c|}{  0.37  } & $\eta_\s{Q} \, \eta_\s{DW} \, \eta_\s{BR}$ \\
    \hline
    \hline
    \multicolumn{5}{|c|}{\textbf{Cavity-Emitter Parameters}} \\
    \hline
    \hline
    Cooperativity & $C$ & $20.6 \pm 1.1$ & $4.4 \pm 0.2$ & \refsub{fig:3}{c} and \refsub{fig:deice-two}{e} \\
    \hline
    Coherent Cooperativity & $C_\s{coh}$ & $8.3 \pm 1.2$ & $1.6 \pm 0.2$ & Figs. \refsub{fig:4}{d} and \refsub{fig:deice-two}{c} \\
    \hline
    Cavity-Emitter Coupling  & $g$ & $(1.94 \pm 0.08)$\,GHz & $(0.99 \pm 0.07)$\,GHz & $\sqrt{C_\s{coh} \kappa \gamma / 4}$ \\
    \hline
    Cavity-Emitter Coupling  & $g^\prime$ & $(1.90 \pm 0.06)$\,GHz & $(1.22 \pm 0.01)$\,GHz & $\sqrt{C \kappa^\prime \gamma_0 / 4}$ \\
    \hline
    Beta Factor & $\beta$ & $0.950 \pm 0.002$ & $0.820 \pm 0.005$ & $\frac{C}{C+1}$ \\
    \hline
    Beta External & $\beta_e$ & $0.199 \pm 0.003$ & $0.254 \pm 0.002$ & $ \frac{\kappa_e}{\kappa} \, \frac{C}{C+1}$\\
    \hline
\end{tabular}
\end{table*}

To measure the loss rate into the waveguide 
$\kappa_e$, we perform reflection measurements.
The reflectivity of a bare cavity can be described by 
\begin{equation}
    R(\Delta_c) = 
    \left|
    1 - \frac{\kappa_e}{\kappa/2 + i\Delta_c}
    \right|^2,
\end{equation}
where 
$\kappa$
is the total loss rate and
$\Delta_c = \omega - \omega_c$
is the detuning between the probe laser and the cavity.
From Section 
\ref{sec:cavity-simulations}
we know that the cavities are under-coupled with 
$\kappa_e < \kappa/2$.
Fitting the cavity reflection under this constraint gives 
$\kappa_e / \kappa = 0.21$ for Device One and 
$\kappa_e / \kappa = 0.31$ for Device Two.
Due to the lensed fiber configuration, we have significant reflections off the fiber tip
\cite{pasini_nonlinear_2024}.
To remove this effect from the measurement, we fit 
$R'(\Delta_c) = A \, R(\Delta_c) + c$,
where 
$A$ is an amplitude and
$c$ is a constant background that is measured with the fiber hovering above the device.
We measured 
$c = \SI{21.9}{\kilo\hertz}$ for Device One and
$c = \SI{20.4}{\kilo\hertz}$ for Device Two.
The maximum reflection counts were $\sim \SI{350}{\kilo\hertz}$ for Device One and $\sim \SI{750}{\kilo\hertz}$ for Device Two.

\section{C-QED Parameters}
\label{sec:cqed_parameters}
In Table \ref{tab:cqed} we gather all the relevant parameters for both of the investigated devices together with their source. Note that the quality factors of both devices varied with time. Importantly, the quality factors extracted from lifetime measurements are not the same as the ones measured in the resonant linescans, since they were performed at different times for both devices. On the other hand, the cavity-emitter coupling $g$ is a parameter that remains largely unaffected by a varying quality factor. In Table \ref{tab:cqed} we show each of the two values of $g$ independently extracted from the lifetime and resonant measurements. We see a match for Device 1 but not for Device 2. The green pulsed laser backtuned the cavity during the lifetime measurements from Fig. \refsub{fig:deice-two}{d,e}, which could explain the bad fit and the mismatch in the value of $g$.

\section{Device Two Data}
\label{sec:device-two}
\begin{figure*}[ht]
    \centering
    \includegraphics[width=\linewidth]{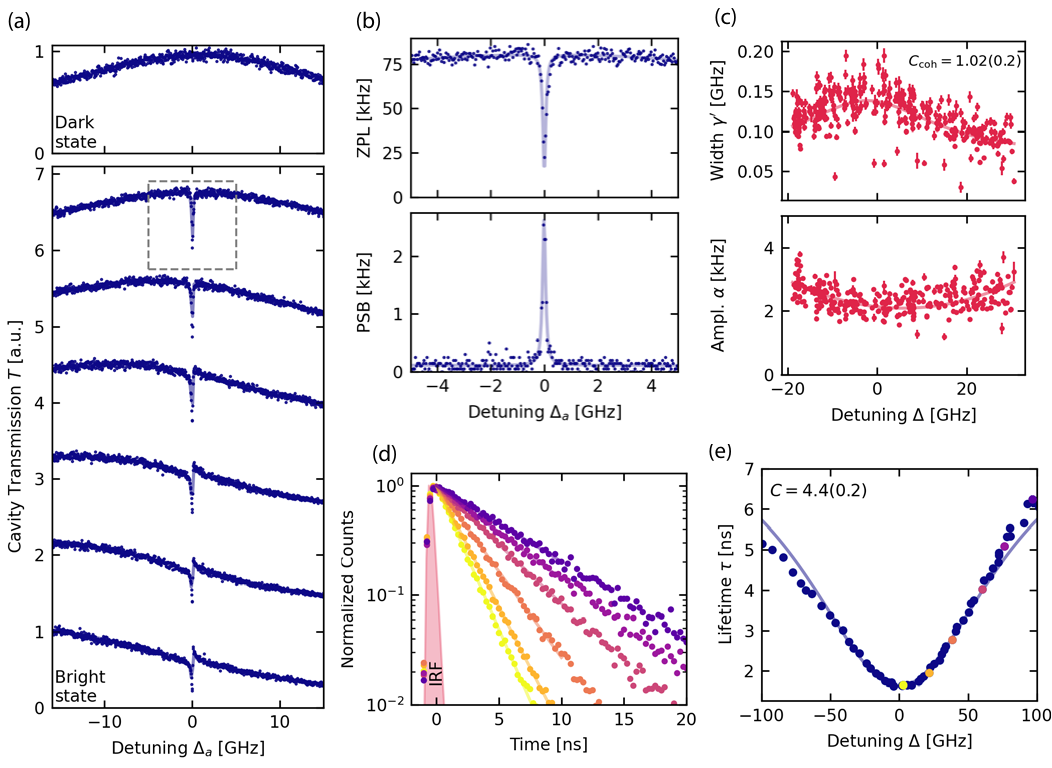}
    \caption{
    Device Two C-QED characterization.
    \textbf{(a-c)}
    Cavity transmission measurements. 
    In (a), the transmission is shown for six different detunings 
    $\Delta \in [17.2, \, 15.7, \, 12.5, \, 8.9, \, 5.1, \, 0.1] \, \si{\giga\hertz}$.
    For explanation, see Fig. \ref{fig:4}.
    \textbf{(d-e)}
    Lifetime measurements.
    For explanation, see Fig. \ref{fig:3}.
    }
    \label{fig:deice-two}
    
\end{figure*}

We have performed the full set of C-QED characterizations for a second device.
This device has only one mirror hole in the outcoupling mirror region compared to four mirror holes for device one.
Based on the simulations of Supplement Section \ref{sec:cavity-simulations}, we expect device two to have a higher outcoupling rate $\kappa_e$, comparable intrinsic losses $\kappa_i$ and thus a lower overall quality factor.
The reflection measurement (Fig. \ref{fig:reflection}) shows that indeed the outcoupling rate is slightly higher 
(\SI{8.2}{\giga\hertz} vs. \SI{5.8}{\giga\hertz}).
However, the intrinsic losses are also significantly higher
(\SI{13.2}{\giga\hertz} vs. \SI{41.8}{\giga\hertz}).
Figures \refsub{fig:deice-two}{a-c} show the cavity transmission measurements and Figures \refsub{fig:deice-two}{d-e} show the lifetime measurements. 
The C-QED parameters extracted from the transmission measurements are $\{g, \kappa, \gamma\}/2\pi = \{\SI{0.904\pm 0.009},\, \SI{36\pm 1.3}, \,\SI{0.052}\}$~GHz.